# Spin-glass behavior in single crystals of hetero-metallic magnetic warwickites $MgFeBO_4$, $Mg_{0.5}Co_{0.5}FeBO_4$, and $CoFeBO_4$

**A. Arauzo[1*], N.V. Kazak[2], N.B. Ivanova[3], M.S. Platunov[2], Yu.V. Knyazev[3], O.A. Bayukov[2], L.N. Bezmaternykh[2], I.S. Lyubutin[4], K.V. Frolov[4], S.G. Ovchinnikov[2,3,5], and J. Bartolomé[6]**

[1]*Servicio de Medidas Físicas. Universidad de Zaragoza, Pedro Cerbuna 12, 50009 Zaragoza, Spain.*

[2]*L.V. Kirensky Institute of Physics, SB of RAS, 660036 Krasnoyarsk, Russia*

[3]*Siberian Federal University, 660074 Krasnoyarsk, Russia*

[4]*Shubnikov Institute of Crystallography, RAS, 119333, Moscow, Russia*
[5]*Siberian State Aerospace University, 660014 Krasnoyarsk, Russia*

[6]*Instituto de Ciencia de Materiales de Aragón. CSIC-Universidad de Zaragoza and Departamento de Física de la Materia Condensada. 50009 Zaragoza, Spain*

Magnetic properties of heterometallic warwickites $MgFeBO_4$, $Mg_{0.5}Co_{0.5}FeBO_4$, and $CoFeBO_4$ are presented, highlighting the effect of Co substitution on the magnetic properties of these compounds. The analysis of magnetization and heat capacity data has shown that these compounds exhibit a spin-glass transition below $T_{SG}$=10, 20 and 22 K, respectively. Using zero field ac susceptibility as entanglement witness we find that the low dimensional magnetic behavior above $T_{SG}$ show quantum entanglement behavior $\chi(T) \propto T^{-\alpha(T)}$ up to $T_E \approx 130$K. The $\alpha$ parameters have been deduced as a function of temperature and Co, indicating the existence of random singlet phase in this temperature region. Above $T_E$ the paramagnetism is interpreted in terms of non-entangled spins giving rise to Curie-Weiss paramagnetism. The different intra- and inter-ribbon exchange interaction pathways have been calculated within a simple indirect coupling model. It is determined that the triangular motifs in the warwickite structure, together with the competing interactions, induce frustration. The spin-glass character is explained in terms of the substitutional disorder of the Mg, Fe and Co atoms at the two available crystallographic sites, and the frustration induced by the competing interactions. The Co substitution induces uniaxial anisotropy along the *b* axis, increases the absolute magnetization and increases the spin-glass freezing temperature. The entanglement behavior is supported in the intermediate phase irrespective of the introduction of anisotropy by the Co substitution.

## 1. INTRODUCTION

Warwickites are mixed borates with general formula $M^{2+}M'^{3+}OBO_4$ which are crystallized in monoclinic or orthorhombic structure. The crystal structure can be represented as the assembly of linear substructures, similar to ribbons, extending along the *c*-axis. The ribbons are formed by four columns of edge - sharing oxygen octahedra at the center of which the divalent and trivalent metallic ions are located (see Fig. 1). There are two crystallographic nonequivalent positions M1 and M2 for magnetic ions. The warwickites are naturally disordered materials since each metal crystalline site may be occupied by any one of the two metals. This disorder generates a broad spectrum of intensities for the exchange and superexchange interactions between the magnetic ions. In highly anisotropic borates, such spectrum yields to disordered quantum magnetic chain type of behavior [1].

The warwickites can be formed with most of the transition metals, allowing for systematic investigations of their physical properties. At present there are reports on only two homo-metallic (M = M') warwickites: $Fe_2BO_4$[2][3],[4],[5],[6] and $Mn_2BO_4$[4],[6],[7], exhibiting both long-range magnetic order. Several studies have been done on different magnetic properties of heterometallic (M ≠ M') warwickites with only one magnetic ion, $MgTiBO_4$ [8],[9],[10], $MgCrBO_4$ [11], $MgFeBO_4$ [11],[12], $NiScBO_4$ [11], $MnScBO_4$ [11], $MgVBO_4$ [11],[13].

At sufficiently high temperature, heterometallic warwickites with just one magnetic metal are paramagnetic and obey the Curie-Weiss law with antiferromagnetic exchange interaction between nearest neighbors. As temperature is lowered, short range interaction within the ribbons gives rise to quasi one-dimensional interactions since $k_BT$ becomes of the order of the intra-ribbon exchange energy. In this temperature range these materials can be described in terms of the Random Exchange Heisenberg AF Chains (REHAC) approximation[14]. At lower temperature interactions there is a 3-dimensional spin-glass transition at $T_{SG}$ when $k_BT$ becomes lower than inter-ribbon exchange interaction [1].

The number of works of hetero-metallic warwickites where both ions are magnetic is extremely small. The crystal structure of $CoCrBO_4$, $NiFeBO_4$, $CoFeBO_4$, and $MnFeBO_4$ have been determined previously [15]. Magnetic properties have been shortly addressed in the case of $NiFeBO_4$, $CuFeBO_4$ and $CoFeBO_4$ [16]. In more recent studies of $Fe_{1.91}V_{0.09}BO_4$ [17],[18], it has been shown that he introduction of Vanadium as a partial substitution of Fe does not alter magnetic properties radically. Indeed, although V acts so as to hinder inter - ribbon Fe – Fe interactions, magnetic ordering also takes place, although at a lower temperature.

TABLE I. Magnetic properties of the warwickites.

| | $T_{\rm ord}$ (K) | $T_{SG}$ (K) | $\theta$ (K) | $\lvert\theta\rvert / T_{SG}$ | Valence, $S$ | Reference |
|---|---|---|---|---|---|---|
| $Fe_2BO_4$ | 155 | | | | $Fe^{2+}$, $S$=2 | 2 |
| | | | | | $Fe^{3+}$, $S$=5/2 | |
| $Fe_{1.91}V_{0.09}BO_4$ | 130 | | | | $Fe^{2+}$, $S$=2 | 17 |
| | | | | | $Fe^{3+}$, $S$=5/2 | |
| | | | | | $V^{2+}$, $S$=3/2 | |
| $NiFeBO_4$ | | 12 | -450 | 37.5 | $Ni^{2+}$, $S$=1 | 16 |
| | | | | | $Fe^{3+}$, $S$=5/2 | |
| $CuFeBO_4$ | | 12 | -200 | 16.7 | $Cu^{2+}$, $S$=1 | 16 |
| | | | | | $Fe^{3+}$, $S$=5/2 | |
| $CoFeBO_4$ | | 30 | -290 | 9.7 | $Co^{2+}$, $S$=3/2 | 16 |
| | | | | | $Fe^{3+}$, $S$=5/2 | |
| $MgFeBO_4$ | | 11 | -278 | 25.3 | $Fe^{3+}$, $S$=5/2 | 11 |
| $MgVBO_4$ | | 6 | -50 | 8.3 | $V^{3+}$, $S$=1 | 11 |
| $MgCrBO_4$ | | 6.5 | -20 | 3.07 | $Cr^{3+}$, $S$=3/2 | 11 |
| $NiScBO_4$ | | 6 | -16 | 2.7 | $Ni^{2+}$, $S$=1 | 11 |
| $MnScBO_4$ | | 2.7 | -60 | 22.2 | $Mn^{2+}$, $S$=5/2 | 11 |
| $MgTiBO_4$ | | | -73 | | $Ti^{3+}$, $S$=1/2 | 11 |
| $Mn_2BO_4$ | 26 | | | | $Mn^{2+}$, $S$=5/2 | 7 |
| | | | | | $Mn^{3+}$, $S$=2 | |

Spin-glass behavior has been reported in a majority of warwickites, showing a relatively low temperature spin-glass transition $T_{SG}$ (Table I). From previous works in homometallic and heterometallic warwickites we may infer that the introduction of a different metal center has the effect of hampering magnetic order, irrespective of this ion being magnetic or not. In the case of heterometallic warwickites with Fe, MgFe, NiFe and CuFe, very close spin-glass transition temperatures are observed ($T_{\rm SG}$ = 11, 12 and 12 K, respectively). This is in contrast with CoFe warwickite (see Table I) where $T_{\rm SG}$ = 30 K. Thus, the effect of introduction of a magnetic ion in addition to Fe has no effect, with the exception of the Co substitution.

The 3-D spin-glass transition temperature $T_{SG}$ shows a frequency dependence that can be described in terms of the dynamical scaling theory with a critical exponent zν [19]. Moreover, magnetic relaxation behavior at $T<T_{SG}$ also shows spin-glass behavior.

These materials have a renewed interest since they can be used as solid state examples of quantum entanglement. Indeed, the intra-ribbon interactions in these systems support the existence of random magnetic chains. In the low-dimensional REHAC region the magnetic susceptibility, used as an entanglement witness [20], proves that the studied compounds can be described as a chain of entangled spins. Therefore, warwickites are good candidates for the experimental study of thermal entanglement and the relation of entanglement with the spin-glass state.

The nature of the low energy phases found in warwickites can be of different types although the most common picture in high disordered warwickite compounds is a random singlet phase (RSP). In the RSP phase spins are coupled in pairs over arbitrary distances. In the renormalization group approach, this random singlet phase is governed by an infinite randomness fixed point. When the amount of disorder decreases, there is a Griffiths phase which emerges, characterized by exponents which depend on the distance to the infinite randomness fixed point [21],[22]. The temperature dependence of the susceptibility follows a power law with a temperature dependent exponent α(T), which allows classifying the behavior as that of a random singlet phase (RSP) in the case of $MgTiBO_4$, and as a Griffiths phase in the pyroborate $MgMnB_2O_6$ [10].

In a recent Mössbauer spectroscopic study as a function of temperature [23] we have also found spin-glass behavior in $MgFeBO_4$ and $CoFeBO_4$. The increased magnetocrystaline anisotropy by Co substitution increases the magnetic viscosity of the magnetic lattice, by freezing magnetic fluctuations below $T_{SG}$.

Within this context, we aim in this work to study the effect of a highly anisotropic magnetic ion, such as Co, in the entangled and in the spin-glass phases of the heterometallic warwickites. We have selected the series $MgFeBO_4$, $Mg_{0.5}Co_{0.5}FeBO_4$ and $CoFeBO_4$, where Co is partially or totally introduced replacing non-magnetic Mg ion.

The paper is organized as follows. In section 2, the structure is described, and in section 3 the experimental procedures are outlined. In section 4 the results of the magnetic characterization of hetero-metallic Mg-Fe, Mg-Co-Fe and Co-Fe warwickites are presented. First, the temperature dependence of magnetization is introduced where the spin-glass transition is clearly manifested in the three compounds. The study of the anisotropy observed in the spin-glass is further analyzed in the following part. Additional relaxation experiments are given as a complementary manifestation of the spin-glass behaviour. Then, ac susceptibility experiments allow analyzing the spin-glass transition within the Dynamical Scaling Theory [19]. Additionally, susceptibility is used as an Entanglement Witnesses in these compounds and the presence of Random Singlet Phase is outlined. In section 5 a superexchange model is given to explain the pertinence of the random exchange antiferromagnetic exchange model in the intermediate phase and the existence of frustration in the spin-glass phase. Discussion of experimental results is made in section 6 and a summary of our conclusions is presented in section 7. In the Supplementary Material (SM) we provide additional crystal structure data and supporting information for the exchange model. SM also contains results of heat capacity measurements.

## 2. STRUCTURAL DETAILS

Detailed crystal data for Mg-Fe, Mg-Co-Fe and Co-Fe warwickites are obtained in a previous work [24] and summarized in Tables SMI and SMII of Supplementary Material [25]. The general features of the crystal structure are typical for warwickites [26]. The metal ions are surrounded by oxygen octahedra. These octahedra are linked by edge sharing and form four - octahedra flat ribbons extending along the *c* - axis (Fig. 1). The row consisting of four octahedra adjoined in the sequence 2 – 1 – 1 - 2 is located across the ribbon. The coordination octahedra around the M2 position form the outer columns of the ribbon and the octahedra around the M1 position form the inner two columns (Fig. SMI (*a)*). The planar trigonal borate group ($BO_3$) located in the voids between the ribbons are attached to them by corner sharing (Fig. SMI (*b)*).

From the structural study on Mg-Fe, Mg-Co-Fe and Co-Fe warwickites [24] it may be inferred that Co and Mg enter into the warwickite structure with divalent state, and Fe with trivalent state. Both (M1 and M2) positions are occupied by a mixture of Mg, Co and Fe atoms, although trivalent Fe ions prefer smaller octahedra: $M1O_6$ in the Mg-Fe and Mg-Co-Fe warwickite, and $M2O_6$ one in the Co-Fe compound [24][23].

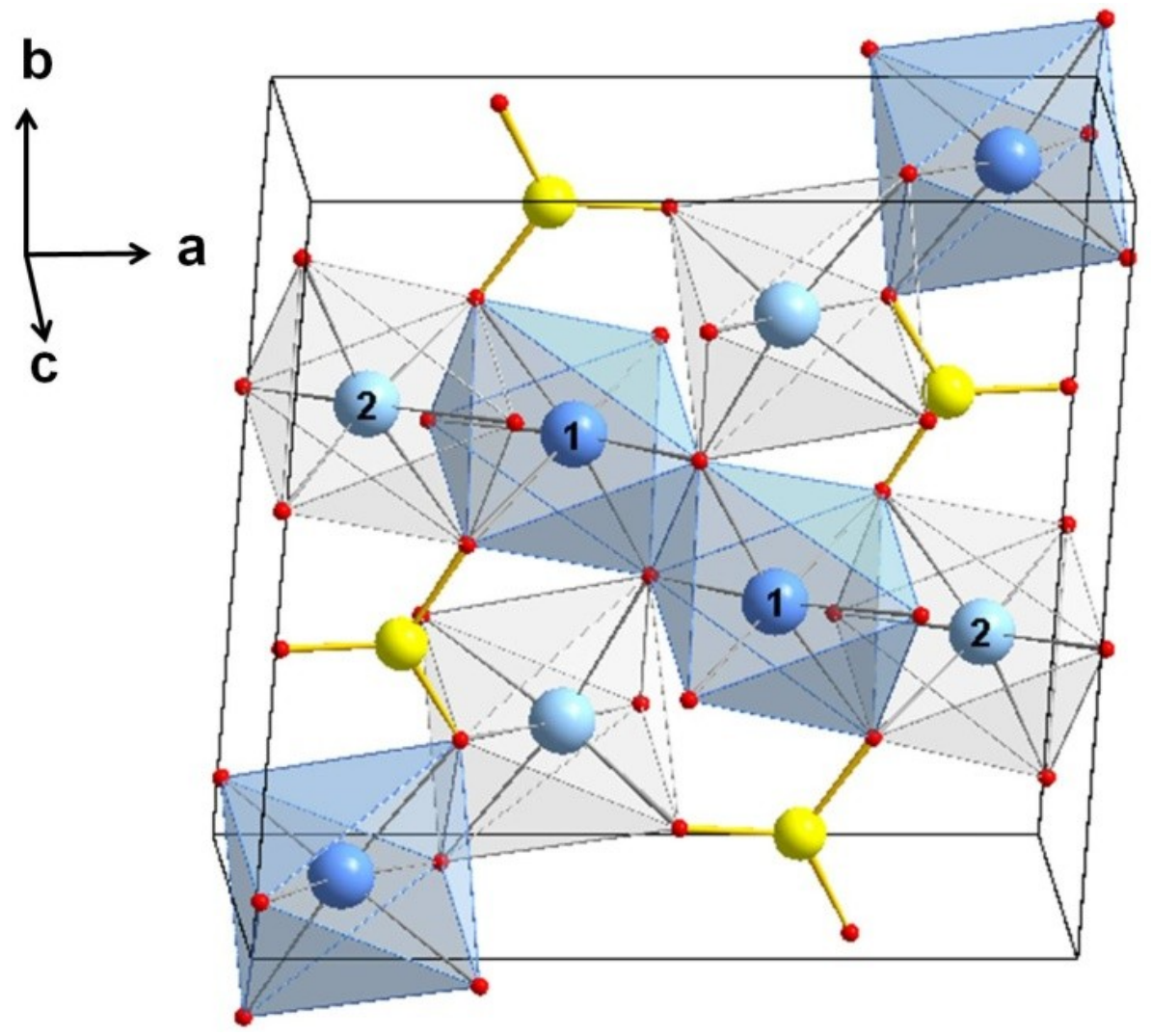

FIG. 1. The schematic structure of the warwickite. The metal cations have octahedral coordination, where the octahedra sharing edges form ribbons. Coordination octahedra around the M1 position (labeled 1) are dark and those around the M2 position (labeled 2) are light. The boron atom positions drawn as yellow circles have trigonal coordination. The sides of the unit cell are shown.

## 3. EXPERIMENTAL PROCEDURE

Single crystals of Mg-Fe, Mg-Co-Fe and Co-Fe, warwickites were grown by the flux method in the system $Bi_2Mo_3O_{12}$ - $B_2O_3$ – CoO – MgO – $Fe_2O_3$ [24]. Needle shape black crystals with a typical size of 0.5 x 0.2 x 5.0 $mm^3$ were obtained.

Ac susceptibility measurements were performed in a superconducting quantum interference device (SQUID) magnetometer with *ac* option, in the frequency range 0.01 < $f$ < 1400 Hz, with an exciting field of 4 Oe. Angle dependent magnetization $M(\theta_H,T)$ on oriented single crystals was measured with a rotating sample holder option in the SQUID magnetometer up to 50 kOe and with a vibrating sample magnetometer up to a bias field of 140 kOe.

Anisotropic samples have been oriented with a four-circle X-ray diffractometer and placed in the sample holder along the desired axis.

Heat capacity as a function of temperature and magnetic field, was measured on single crystals using a Quantum Design PPMS (Physical Properties Measurement System). The crystals were glued to the sample holder with Apiezon grease.

## 4. MAGNETIC PROPERTIES

In this section a thorough study of magnetic properties of the three compounds has been carried out. The analysis of the spin-glass transition and the study of the entangled phases, in relation to the introduction of the Co magnetic ion in the $MgFeBO_4$ compound have been the main subjects of analysis. Most of the measurements are carried out on single crystals, where special emphasis is done in studying the influence of the anisotropy of the different magnetic phases.

In the case of $MgFeBO_4$, there exist some previous results where the spin-glass transition is observed at $T_{SG}$ = 11 K [11]. At higher temperatures a Curie-Weiss law is obeyed, with a negative intercept indicative of AF interactions ($\theta_N$ = -278 K). As $T$ is further decreased there is a fluctuation regime starting at 100 K below which magnetic susceptibility is described by a power law $\chi \propto T^{\alpha}$, with $\alpha$ = 0.54, characteristic of random exchange Heisenberg AF chain (REHAC). From the $\theta_N$ = -278 K value the AF exchange coupling can be derived, as $\theta_N = 2zJS(S+1)/3$ ($S$ = 5/2 for $Fe^{3+}$), $J/k_B$ = -23 K. An increase of the magnetic susceptibility below $T_{SG}$ is observed.

A short note about magnetic properties of CoFe warwickite is also found in the literature [16]. In that work, a low temperature transition to an antiferromagnetic state with a weak ferromagnetic component is observed at 30 K. We consider this temperature as an indication of a spin-glass transition.

Additionally, in a recent Mössbauer study of $MgFeBO_4$, and $CoFeBO_4$low warwickites, spin-glass behavior is revealed at low temperature, with spin-freezing temperatures $T_{SG}$ of 15.2 and 33.2 K for Mg- and Co- warwickites, respectively [23].

### 4.1. Magnetization temperature dependence

Field cooled (FC) and zero field cooled (ZFC) dc magnetization measurements as a function of temperature were performed on a single crystal with an applied field of 0.5 kOe at different crystal orientations. Results for the three compounds are shown in Fig. 2.

FC - ZFC experiments show the typical spin-glass cusp-like maximum in the ZFC curve with a strong thermo-irreversibility between the FC and ZFC magnetization at temperatures below the maximum and the flattening out of the FC magnetization at low temperatures. Irreversibility is found for the three compounds below a critical temperature that we assign to the Spin-Glass transition temperature $T_{SG}$ = 10, 20 and 22 K in the series, Mg-Fe, Mg-Co-Fe and Co-Fe warwickites, respectively. Note that the $T_{SG}$ is doubled by the introduction of Co. Actually, the $T_{SG}$ is much larger for Co warwickites than for the other reported heterometallic warwickites (see Table I).

It can be observed in Fig. 2(a) that the magnetic anisotropy is negligible for Mg-Fe. In contrast, anisotropy is found in the Mg-Co-Fe warwickite, though it is small (Fig. 2(b)). It points out clearly that the $Co^{2+}$ ion induces this anisotropy. This is somehow to be expected since the $Fe^{3+}$ has no orbital momentum, whereas the $Co^{2+}$ in the low symmetry coordination has an orbital contribution caused by the relevant spin-orbit coupling that gives rise to single ion anisotropy.

The anisotropy is far larger in the Co-Fe compound with respect to the Mg-Co-Fe warwickite. Noteworthy, in the CoFe compound there is a factor three increase in the magnetization for the orientation along $b$ axis with respect to needle direction (Fig. 2(c)). The $c$ axis seems to be a hard magnetization direction, while the easy axis lies along the $b$ direction. The maximum of the ZFC curve is at 22 K in the three orientations. The low $T$ behavior is slightly different when the field is oriented along the hard axis. The FC curve along $a$ or $b$ axis is flattened below $T_{SG}$, which is characteristic of spin-glass behavior, whereas along the $c$ axis, the FC magnetization increases below the transition temperature.

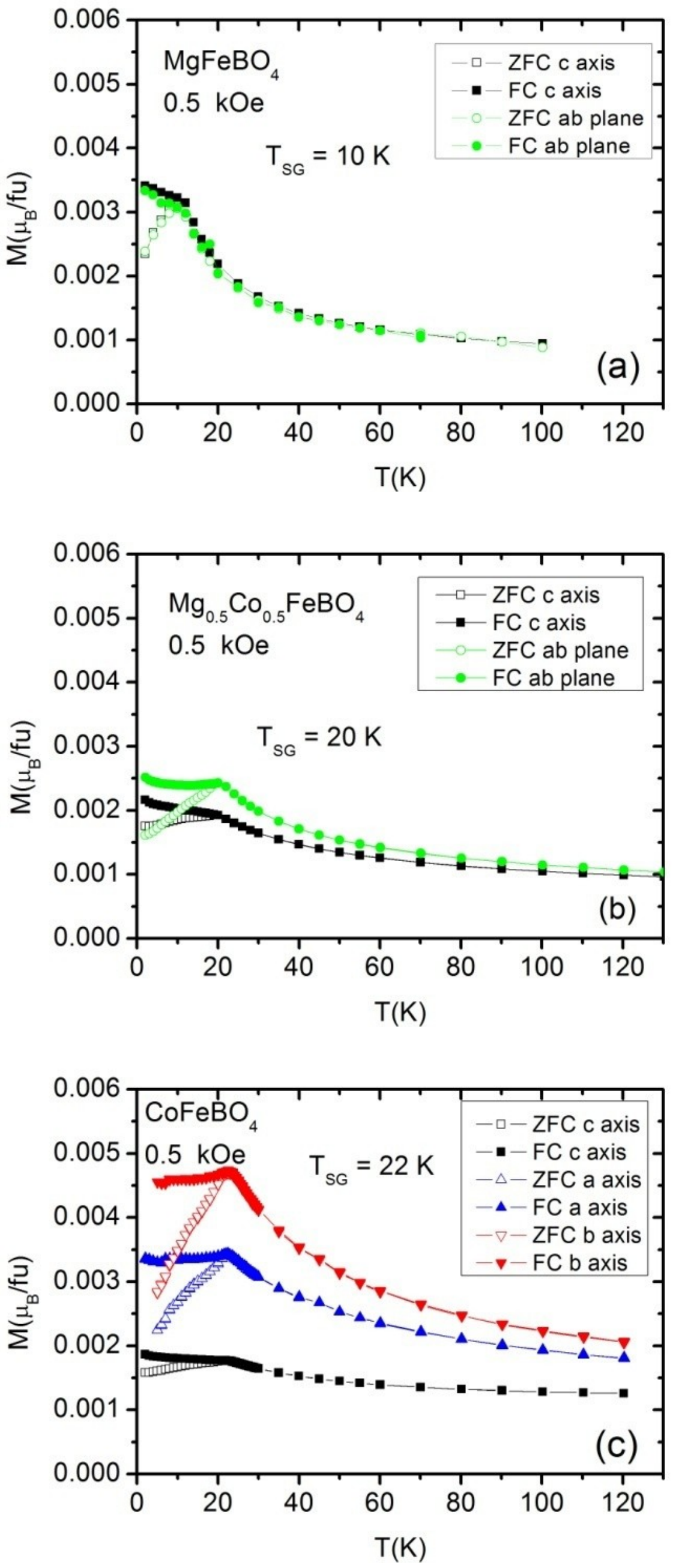

FIG. 2. (Color on line) Magnetization temperature dependence, FC and ZFC curves, showing a spin-glass transition for the studied warwickites. a) MgFe warwickite where no anisotropy is observed; b) MgCoFe warwickite showing small anisotropy and c) CoFe warwickite with well separated curves for the three main axes of the crystal.

The magnetic heat capacity of these samples (see SM, section 3) presents a rounded shape indicative of absence of long range order, and compatible with spin-glass behavior [19].

## 4.2. Magnetic hysteresis

Given the anisotropic behavior observed, a deeper insight can be obtained performing angle dependent magnetization experiments. Indeed, with the rotating sample holder option which allows measuring the projection of the magnetization along the field direction, the easy axis of magnetization as a function of temperature for the two Co compounds can be found. By rotating the sample along a given axis in the presence of an external magnetic field, induced magnetization along the magnetic field direction is measured.

### *4.2.1. $CoFeBO_4$*

This compound exhibits the highest anisotropy. When rotating the sample along the *c* axis, the maximum in the magnetization above $T_{SG}$ is obtained for the field parallel to the *b* axis. The same result is observed when rotating along the *a* axis. Magnetization is maximum when magnetic field is parallel to the *b* axis and minimum at 90º, with a 180º periodicity. Therefore the Easy Magnetization Direction is the *b* axis.

When the external field is relatively low, for $H$ = 0.5 kOe, below $T_{SG}$ the magnetization is maximum at the initial orientation of the crystal after field cooling from $T > T_{SG}$, obtaining the minimum at 180º, independently of the crystal orientation. This behavior indicates that magnetization is frozen and does not rotate with the external magnetic field, thus the measurement just reflects the projection of the invariant thermoremanent magnetization. The observed variation with the rotation angle, $\theta$, can be fitted to a cosine function for $\theta > 100$ º (see Fig. 3a). Therefore, in the spin-glass state, the magnetization does not follow field orientation as we rotate the sample. Instead, the magnetization remains anchored along the FC axis.

When rotating experiments are performed with a high field of 50 kOe, a slightly different behavior is obtained at low temperatures, although fully compatible with the spin-glass character of the material. For $T <$ 10K, a hysteretic behaviour can be observed during the rotation. The magnetization for a field of 50 kOe follows the field direction, but there is an angular shift, which increases as $T$ decreases. The obtained value at $\theta$ = 0 after completion of the whole rotation from $\theta$ = 0 to 360º and back to $\theta$ = 0, is much lower than the initial value (see Fig. 3b).

Some anisotropy remains even at 100 K. Below this temperature, as $T$ decreases magnetization increases up to a maximum value at $T_{SG}$ = 22 K. For lower temperatures the starting magnetization at $\theta$ = 0 is in coincidence with the value at $T_{SG}$. This is one of the characteristics of Spin-glasses, also shown in the FC curves. The minimum value, however, decreases for the lowest temperatures.

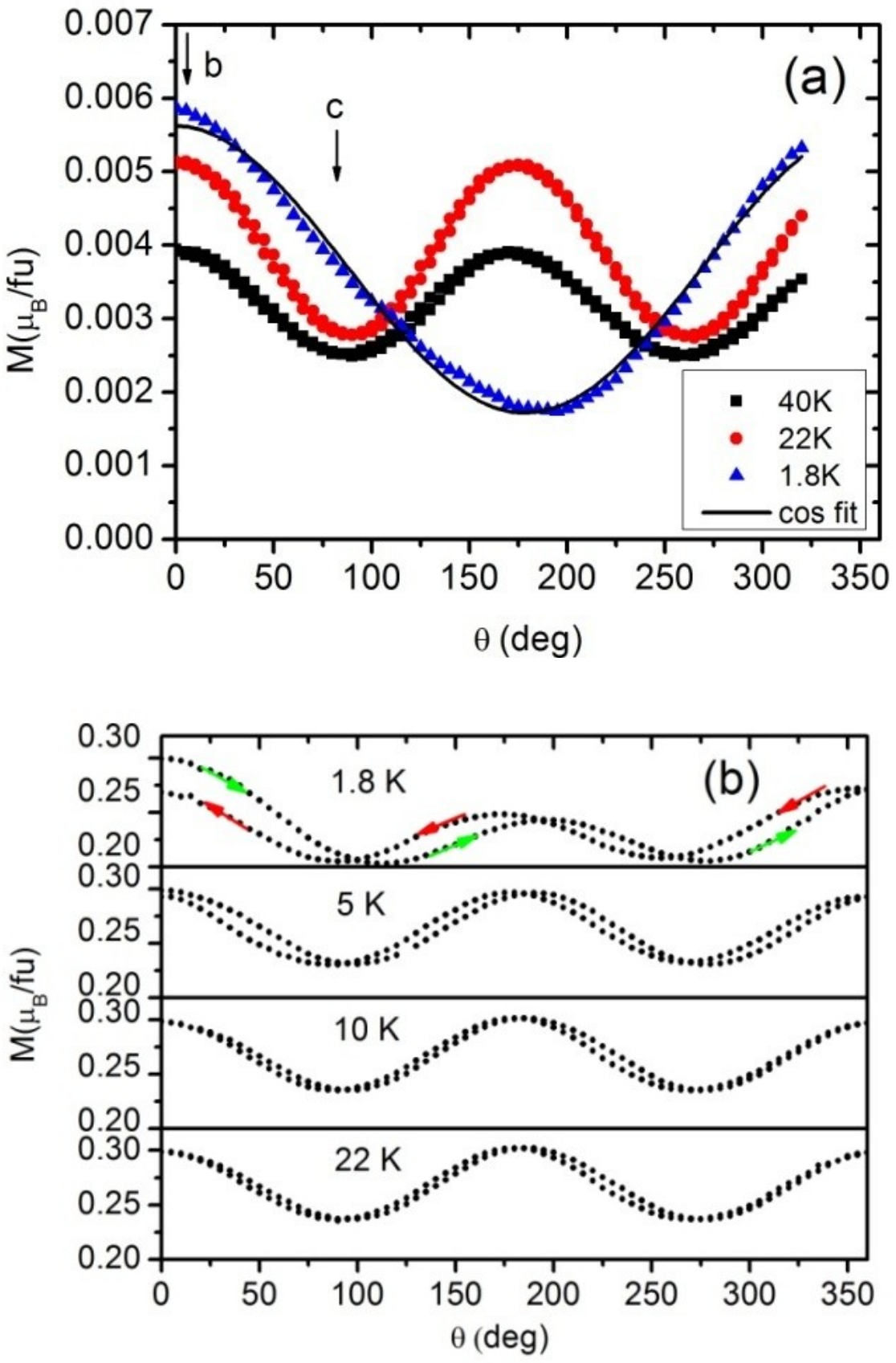

FIG. 3. Magnetization upon rotation for $CoFeBO_4$. a) Rotation around *a*- axis at $H$ = 0.5 kOe. The fit to a cosine function for $T$ = 1.8 K is also shown, where $M = 0.0036 + 0.0019*\cos\theta$ ($\mu_B$ per formula unit). b) Rotation around *c*-axis at $H$ = 50 kOe. Arrows show the rotation scan at 1.8 K, from $\theta$ = 0º to $\theta$ = 360º, green arrows, and back from $\theta$ = 360º to $\theta$ = 0º, red arrows.

### *4.2.2. $Mg_{0.5}Co_{0.5}FeBO_4$*

When the crystal is rotated along the *c* axis, we find that magnetization has a maximum along the *a* axis (Fig. 4) contrary to the Co-Fe warwickite, although below $T_{SG}$ a secondary maximum in the magnetization when the field is aligned along the *b* axis is observed. Anisotropy, even if weak, is noticeable up to high temperatures above $T_{SG}$ and it follows the same trend as in the Co-Fe compound. At 1.8 K the $M(\theta)$ pattern is rather complex due to the high magnetic viscosity at these low temperatures and possibly due to competing anisotropies.

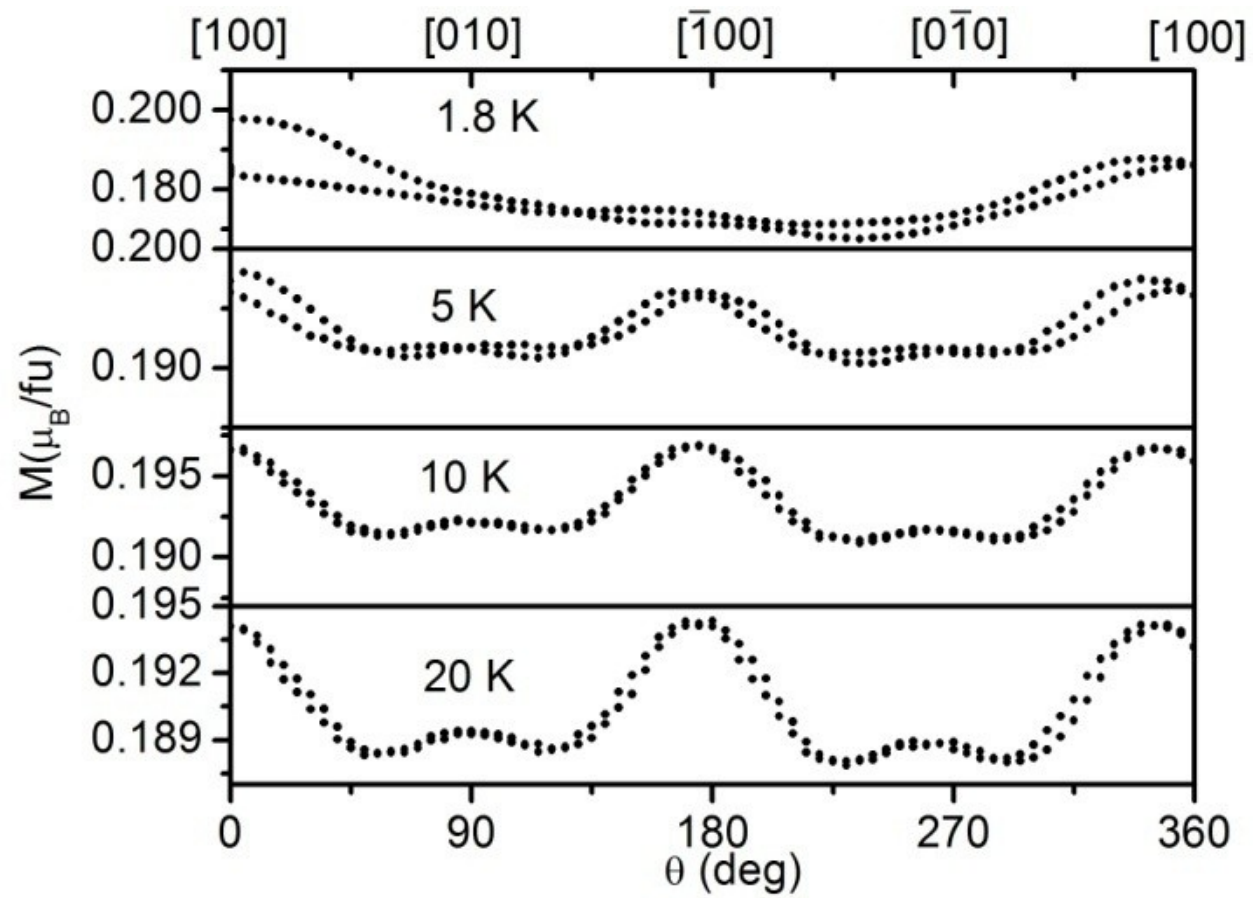

FIG. 4. Magnetization upon rotation around *c* axis for $Mg_{0.5}Co_{0.5}FeBO_4$. $H$ = 50 kOe

A comparison of the anisotropy as a function of temperature for these two compounds is depicted in Fig. 5, where the maximum and the minimum value of the magnetization when rotating along the *c* axis is presented.

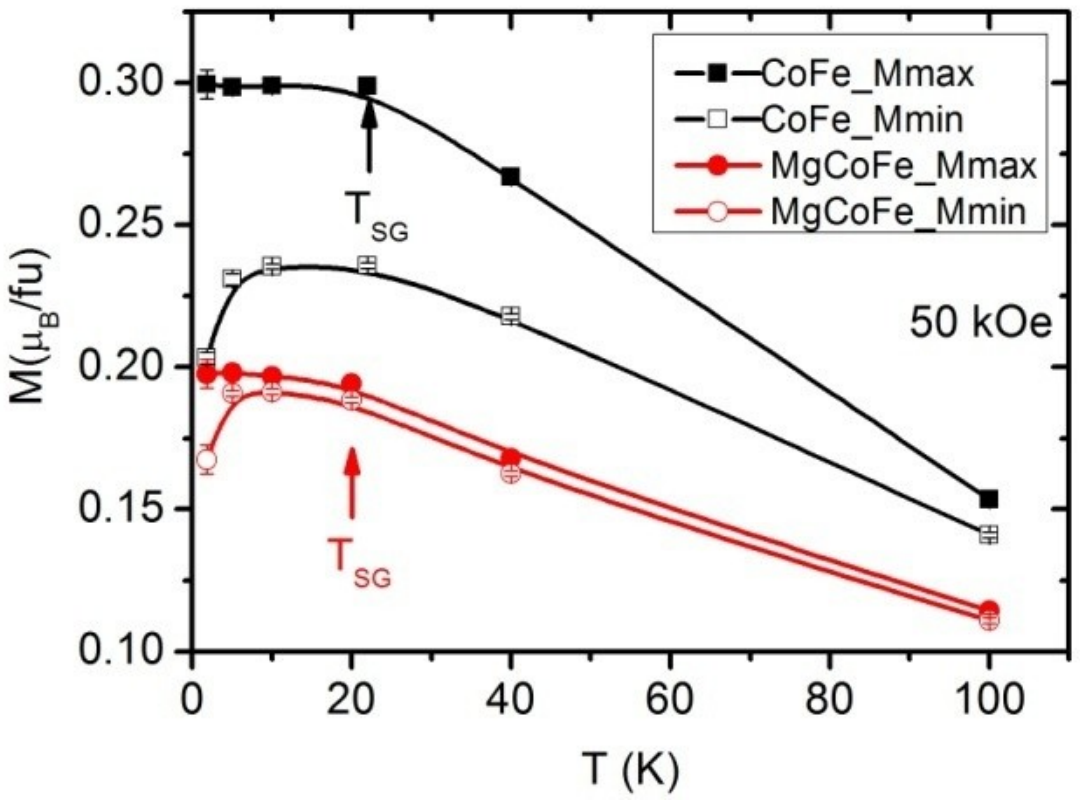

FIG. 5. Extreme magnetization values as a function of temperature upon rotation around *c* axis. $H$ = 50 kOe. Values for $CoFeBO_4$ and $Mg_{0.5}Co_{0.5}FeBO_4$.

Hysteresis loops at low $T$ also show an anisotropic behavior. Remanence and coercive field vary with orientation, being both larger for the easy axis. Nevertheless, for all orientations, a displaced hysteresis loop is observed, which is a signature of the spin-glass state. Hysteresis cycle is recorded after 50 kOe FC from $T > T_{SG}$. This induces thermo remanence (TRM), which is well noticed at $H = 0$ in the *ab* plane (see Fig. 6). As the loop is traced out, this metastable TRM decreases with time, giving a lower value at 50 kOe after the whole cycle is completed. When the loop is traced up to 140 kOe along the easy axis (*b* axis), the hysteresis cycle is symmetric. Therefore in this case TRM at 140 kOe and 2.5 K is, most likely, compensated by the high field. Saturation is never attained even at such a large field as 140 kOe. The hysteresis loop closes at the maximum field, showing no reversibility. Similar results are obtained in the other two compounds, although with lower values of the TRM and coercivity.

Above $T_{SG}$, magnetization can not be fitted to a power law $H^{1-\alpha}$ behavior as found for $MgTiBO_4$ [8], as could be expected for a quantum magnetic chain type of behaviour. Lower $T$ and stronger $H$ conditions would be needed in order to fulfill that power-law dependence [14].

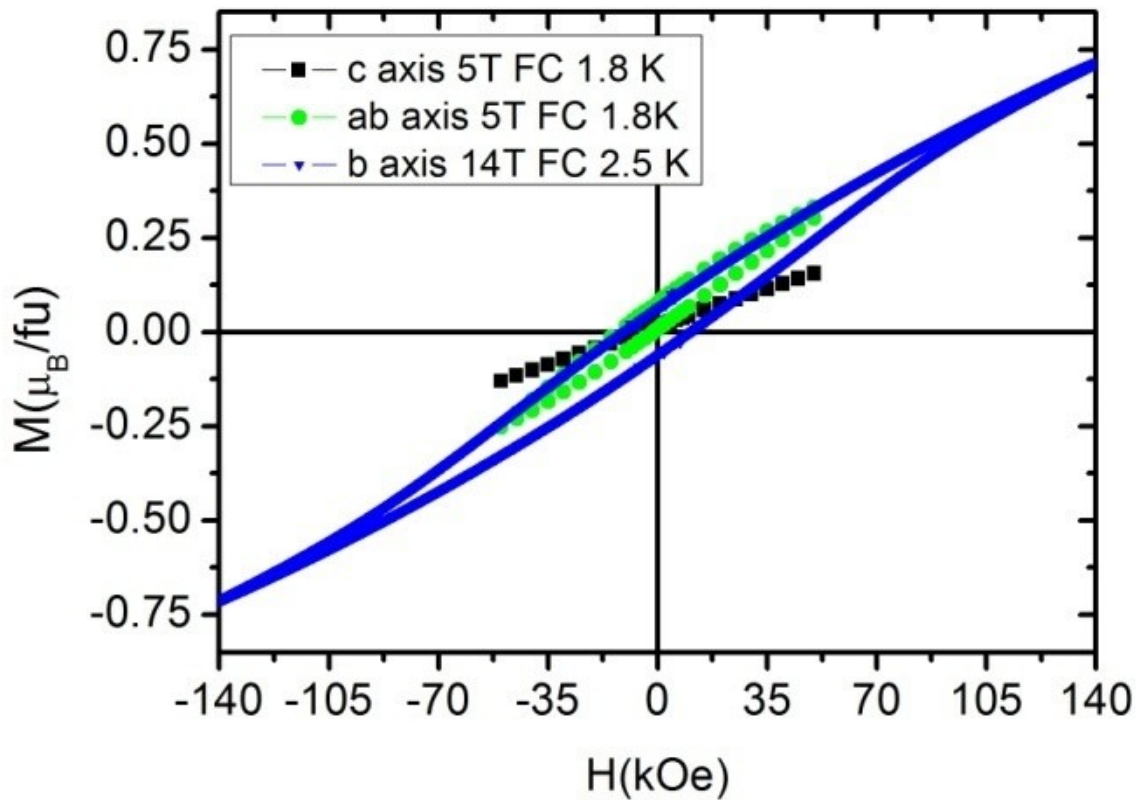

FIG. 6. Hysteresis loops for a $CoFeBO_4$ single crystal at 1.8 K after 50 kOe FC for field parallel (*c* axis) and perpendicular to needle axis (*ab* plane) and after 140 kOe FC at 2.5 K for field parallel to the easy axis.

### 4.3. Magnetic Relaxation.

Magnetic relaxation experiments at low temperature have been performed in $CoFeBO_4$ to characterize the spin-glass behavior. The characteristic features of the glassy nature of the compound at $T < T_{SG}$ are detected.

The relaxation experiments have been carried out by measuring the Low Temperature Field Cooled or Thermoremanent Magnetization (TRM) [19]. In a TRM experiment, the sample is cooled in a weak field, from high $T$ to a $T < T_{SG}$. Then, after a waiting time, $t_w$, the field is set to zero and the magnetization relaxation as a function of time, $M(t)$ is recorded.

TRM of a single crystal oriented parallel to field was measured at 1.8, 10 and 18 K, after FC at 500 Oe from 50 K. $M(t)$ has been measured after $t_w$ = 10 s. Results are shown in Fig. 7.

The obtained $M(t)$ data have been fitted to the sum of a stretched exponential and a logarithmic decay:

$$M(t) = M_0 * \exp(-(t/t_p)^{1-n}) + SH * \ln(t) \qquad (1)$$

Where $M_0$ and $t_p$ depend upon $T$ and $t_w$, $1-n$ is the exponential grade, which goes from $n = 0$, where we have a Debye single time constant exponential relaxation, to $n = 1$, where $M(t)$ would be constant (apart from the logarithmic term). The value of $n$ governs the relaxation rate from very strong to none at all. $SH$ is the relaxation rate constant in dynamical equilibrium, which only weakly depends upon the time and waiting time. The time decay is logarithmic for $t << t_w$ and $t >> t_w$. Results of the fit parameters are summarized in Table II.

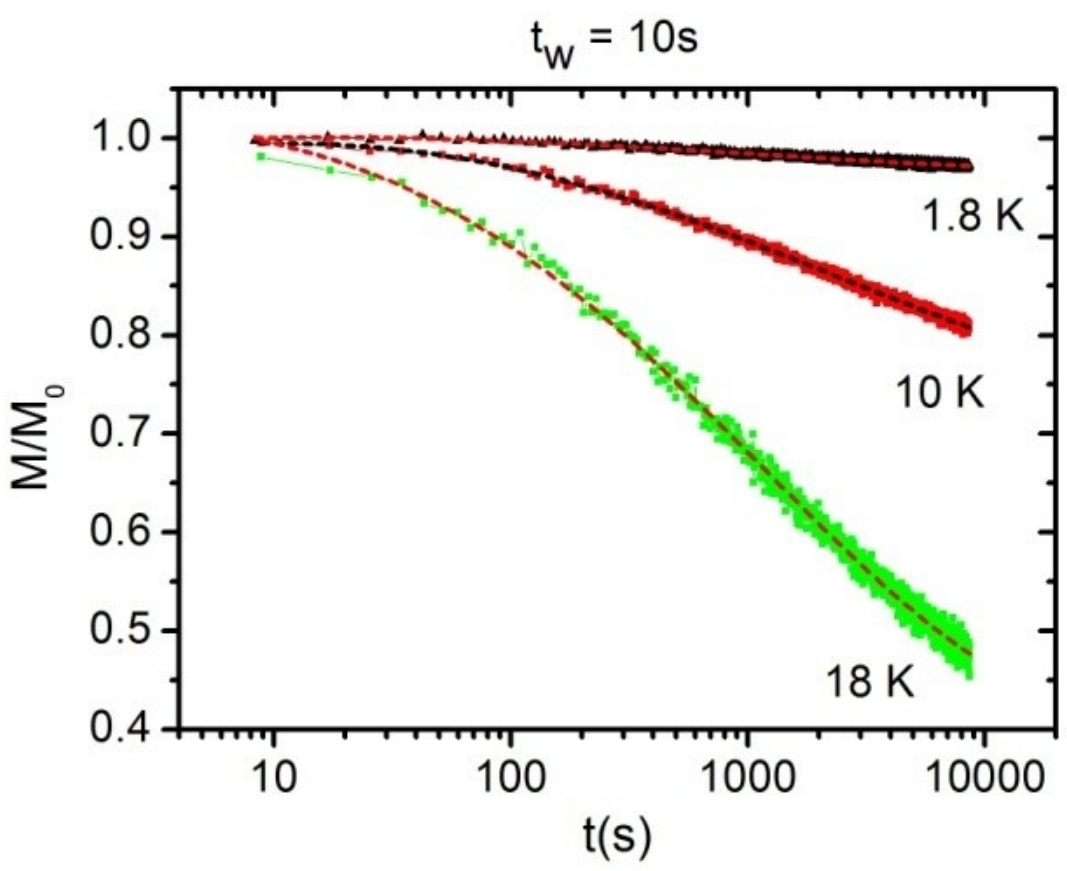

FIG. 7. Magnetic relaxation of $CoFeBO_4$: Normalized TRM (FC at 0.5 kOe) for different temperatures in log - log scale. The fit curves are also shown.

From these relaxation experiments we can see the tendency of the stretched exponential grade *1-n* to decrease as *T* decreases. The relaxation is slowed down at low temperatures.

TABLE II. Fit parameters obtained for *M(t)* as a function of *T*.

| | 1.8 K | | 10 K | | 18 K | |
|---|---|---|---|---|---|---|
| | Value | Standard Error | Value | Standard Error | Value | Standard Error |
| $M_0$ | 1.54 | 0.03 | 1.14 | 0.01 | 1.10 | 0.01 |
| $t_p$ | 1292 | 122 | 2332 | 113 | 1252 | 30 |
| 1-*n* | 0.113 | 0.005 | 0.236 | 0.004 | 0.348 | 0.007 |
| *SH* | 0.058 | 0.002 | 0.057 | 0.001 | 0.036 | 0.001 |

The magnetic relaxation and memory effects give strong evidence of glassy dynamical properties associated with magnetic disorder and frustration.

## 4.4. AC magnetic susceptibility

### *4.4.1. Low T: spin-glass behavior*

The spin-glass transition can be clearly observed in *ac* magnetic susceptibility temperature dependence in the three compounds. For these measurements a single crystal was not large enough to give a good signal to noise ratio, so in most cases the collective signal for several samples was measured, all oriented along the easy plane. As an example, the temperature behavior of real $\chi'$ and imaginary $\chi''$ components of magnetic susceptibility of Mg-Co-Fe warwickite are shown in Fig. 8, where a cusp-like maximum at about 20 K is observed at low frequency. As frequency increases, the maximum shifts slightly but neatly towards higher temperatures, decreasing its intensity. Temperature shift is relatively small for a change in frequency of four decades. The increase of the maximum intensity at low frequencies is about a 5% of the peak value. A similar increase is found for the Co-Fe warwickite, and a 6% in the case of the Mg-Fe compound. The out-of-phase ac susceptibility signal is only plotted for a frequency of 10 Hz, showing a step like transition at $T_{SG}$.

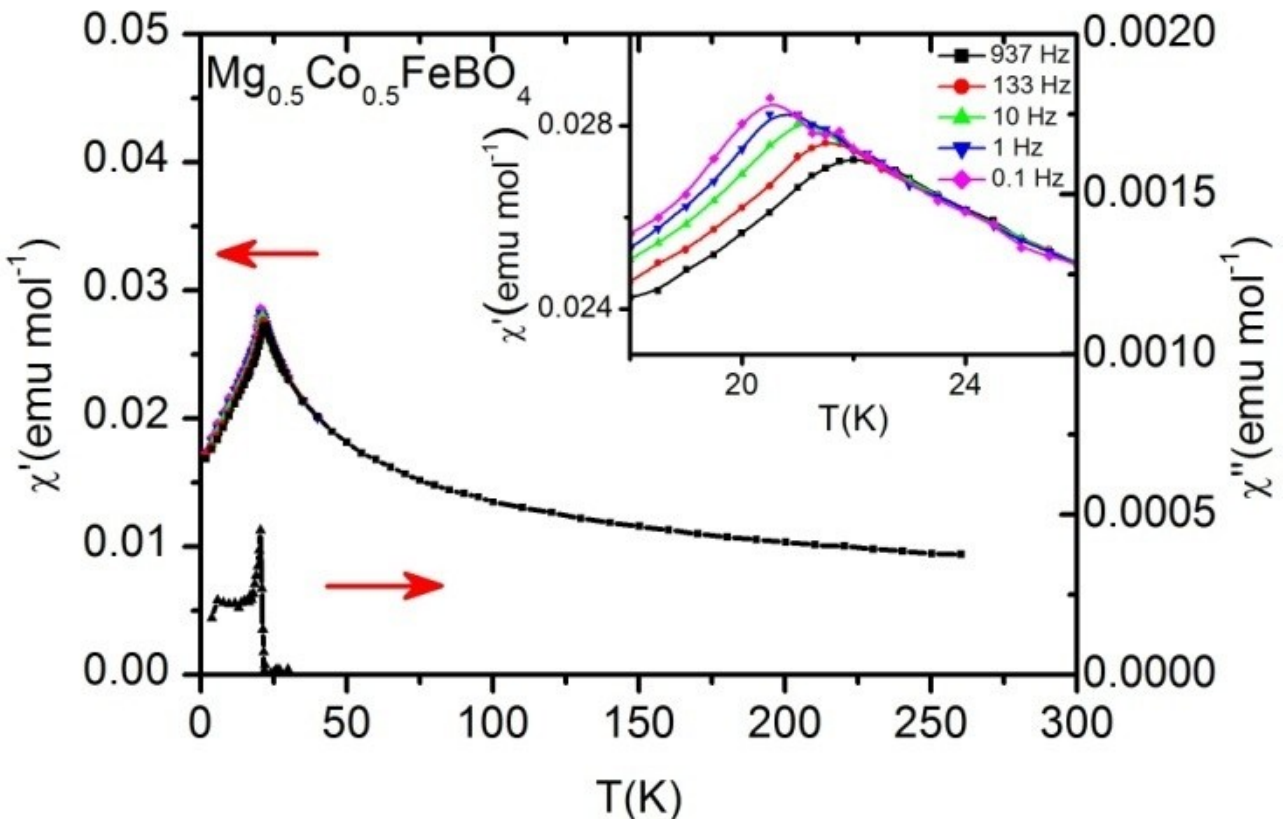

FIG. 8. AC Magnetic susceptibility as a function of temperature and frequency for $Mg_{0.5}Co_{0.5}FeBO_4$. Out of phase component is represented in the secondary axis for 10 Hz. Inset: Larger temperature scale showing the maximum frequency dependence.

The frequency dependence of the $\chi_{ac}$ maximum temperature has a clear spin-glass tendency signature. A way to evaluate the frequency sensibility is to calculate the *p* factor, defined as $p = \Delta T_p /[T_p \Delta(\log f)]$. This value is of about 0.025 for the Co-Fe, 0.021 for the Mg-Co-Fe and 0.014 for the Mg-Fe warwickite, close to values found in canonical spin-glasses where *p* varies in between 0.005 and 0.018 [19]. This low *p* value anticipates the failure of an Arrhenius law fitting, which gives non-physical parameters.

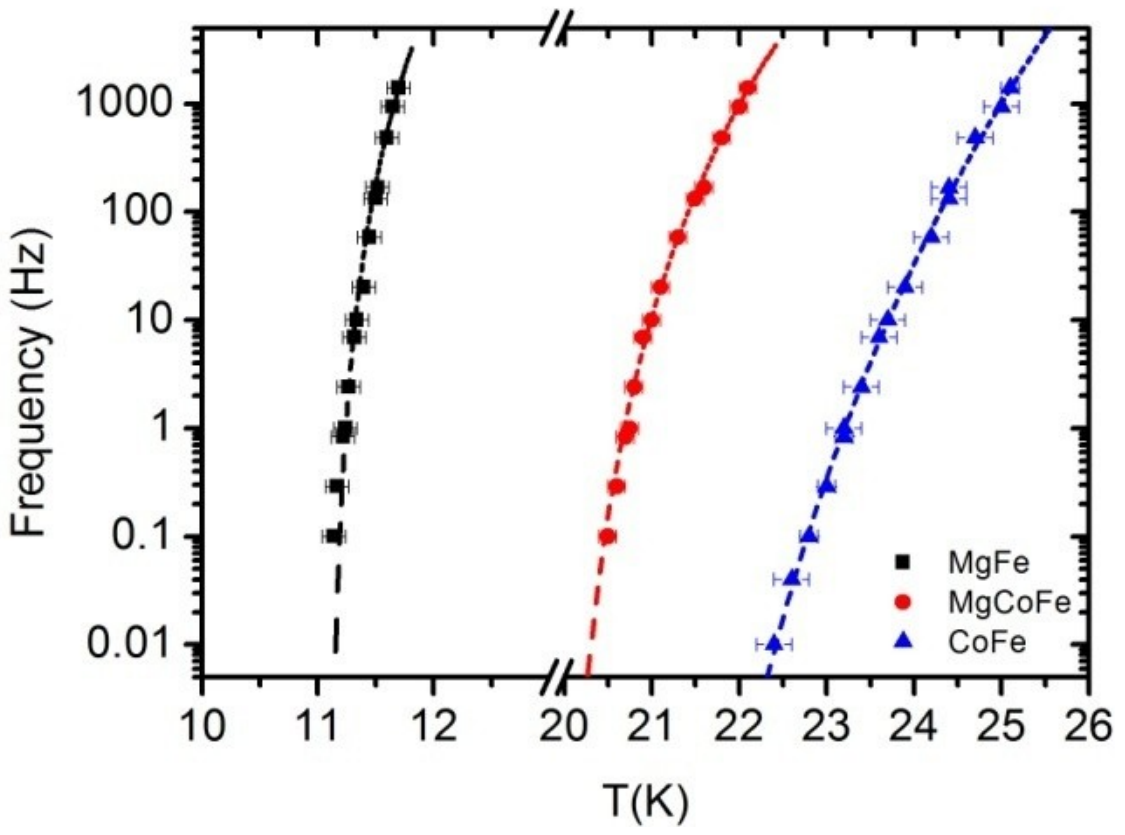

FIG. 9. Variation of the spin-glass transition temperature as a function of frequency. Data obtained from $\chi_{ac}(\omega)$. In red, fit to a critical slowing down law.

Instead, we have made use of the Dynamical scaling theory near a phase transition at $T_c$ to obtain a fit of the maximum frequency dependence (see Fig. 9). According to this hypothesis, the relaxation time close to the transition follows the critical slowing down law, which in terms of frequency stays:

$$f = fo\,(T(\omega)/T_c - 1)^{z\nu} \qquad (2)$$

where $T(\omega)$ is the spin-glass transition temperature as a function of the frequency and $T_c$ is the phase transition temperature in the limit of zero frequency.

The best fit parameters are given in Table III. The spin-glass transition temperature obtained from the FC/ZFC experiments is given for the sake of comparison.

TABLE III. Best fit parameter for the $\chi_{ac}$ maximum frequency dependence.

| | $T_{SG}(K)$ | $T_c(K)$ | $f_o$ (Hz) | $z\nu$ |
|---|---|---|---|---|
| Mg-Fe | 10 | 11.1±0.1 | 3.0±0.1·$10^9$ | 5±1 |
| Mg-Co-Fe | 20 | 19.9±0.2 | 6.7±0.1·$10^9$ | 7±1 |
| Co-Fe | 22 | 20.4±0.2 | 1.2±0.1 $10^{12}$ | 14±1 |

The obtained parameters are quite reasonable for a spin-glass as compared to those found in other systems. Moreover, the $T_c$ values are in good concordance with the experimentally obtained $T_{SG}$. The dynamical critical exponent, $z\nu$, agrees well with those reported for spin-glasses, namely in between 4 and 12 [19].

The $z\nu$ and $f_0$ values increase with increasing Co content, indicating a faster dynamics in the freezing process in CoFe warwickite. This result agrees very well with the tendency obtained for the mean $p$ value calculated for these compounds. Therefore, in these warwickites, a decreasing degree of disorder and frustration takes place upon substitution of Mg by magnetic Co ion.

### 4.4.2. Intermediate T: Random Singlet Phase

In the intermediate $T$ range, in between the spin-glass transition and the paramagnetic behavior, we have a Fluctuation regime where $\chi$ is proportional to $T^{\alpha}$ (characteristic of random exchange Heisenberg AF chain REHAC). We observe such a potential dependence in all the compounds in the log-log $\chi(T)$ plot. The exponent is similar for the pure compounds, $\alpha = 0.62$ and 0.63 for Mg-Fe and Co-Fe respectively, and lower for the mixed warwickite, 0.45. In a previous work [11] they obtain $\alpha = 0.54$ for Mg-Fe warwickite, although this value depends on the fitted temperature range. Similar values of $\alpha$ have been found in the $S$=1/2 $MgTiOBO_3$ warwickite, where a further analysis allows to quantify quantum entanglement in this low-dimensional spin system[10].

In the temperature range where random magnetic chains are formed, magnetic susceptibility can be used as a macroscopic entanglement witness. As demonstrated elsewhere [20], when the condition $\chi < NS/3k_BT$ is fulfilled, where $\chi$ is the averaged zero-field susceptibility, $S$ is the spin of the system and $N$ is the number of spins per mol, the solid state system contains entanglement between individual spins. Entanglement can be measured by the quantity $E$, defined as:

$$E = 1 - k_B T\left(\frac{\chi_x+\chi_y+\chi_z}{(g\mu_B)^2 NS}\right) \qquad (3)$$

According to this definition, the system is entangled when $E > 0$. This parameter quantifies the entanglement, which is maximum, $E$=1, for the extreme case of a singlet state of $N$ spins, where $\chi_x+\chi_y+\chi_z=0$.

We have quantified the entanglement in the three studied warwickites taking the measurement of the magnetic susceptibility for a collection of crystals as a mean value of $\chi_x+\chi_y+\chi_z$. Following the calculation of magnetic susceptibility as a function of the sum of variances of individual spins [20], the contribution of the different $S$=5/2 for the $Fe^{3+}$ and $S$=3/2 for $Co^{2+}$ has been considered as additive in Eq. 3. Therefore for a system with two sets of different spins, $S_1$ and $S_2$, the entanglement witness can be quantified as:

$$E(T) = 1 - k_B T\left(\frac{3\chi_{ac}(T)}{\mu_B^2(g_1^2 N_1 S_1 + g_2^2 N_2 S_2)}\right) \qquad (4)$$

Results are given in Figure 10, where it can be clearly seen that entanglement is present in these systems up to temperatures of about $T_E$=130K, above which a Curie-Weiss paramagnetic behavior is foreseen.

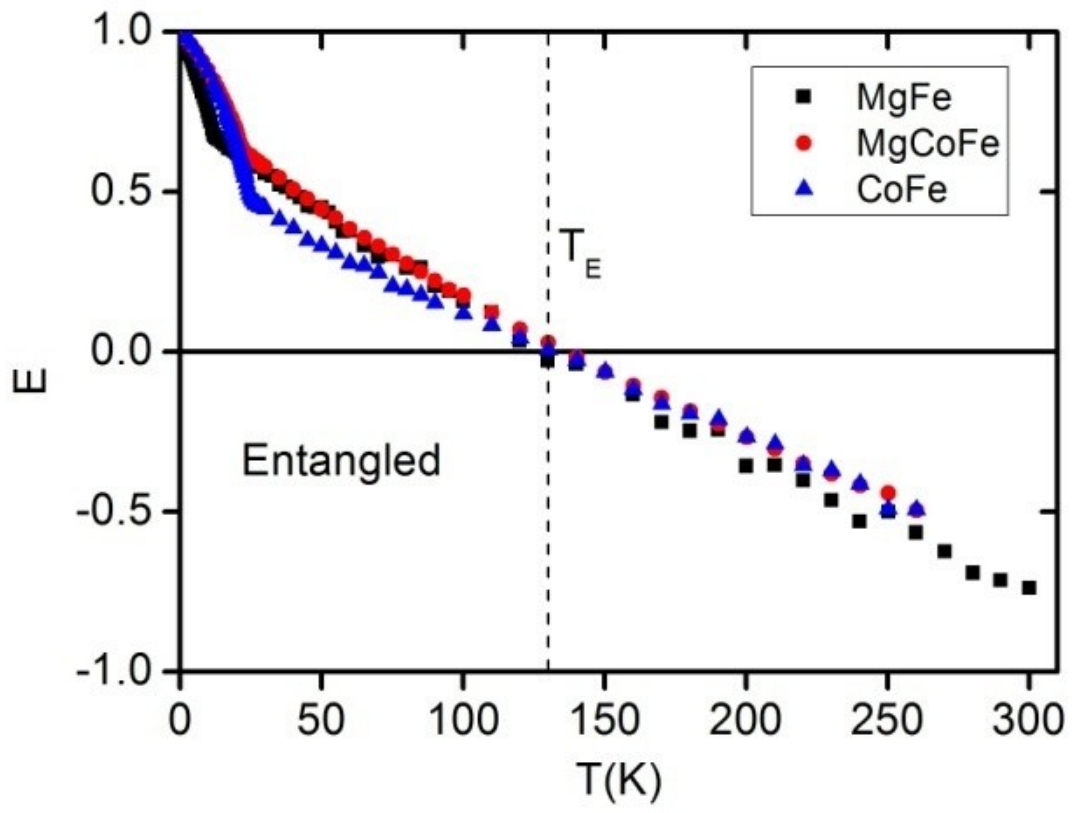

FIG. 10 Calculation of $E$(T) for the three compounds (Eq. 3). Entanglement ($E$>0) is observed for temperatures below 130 K.

On the other hand, the analysis of the temperature dependence of the $\alpha$ exponent gives insight into the phase diagram of the random magnetic chains [10]. We are dealing with $S \geq 1/2$ systems, with $S$=5/2 REHAC for Mg-Fe compound, and $S$=3/2 and $S$=5/2 REHAC system for the Co-Fe and Mg-Co-Fe warwickites. Therefore these systems, with strong disorder are prone to form a Random Singlet Phase (RSP), where singlets of pairs of arbitrarily distant spins are formed [27]. For RSP, experimental magnetic susceptibility can be described with [28]:

$$\chi \propto \frac{1}{T ln^2(\Omega_0/T)} \qquad (5)$$

which is equivalent to a $T^{\alpha(T)}$ function with a slow varying $\alpha(T) = 1-2/ln(\Omega_0/T)$. Magnetic susceptibility data have been fitted to Eq. 5 for $T_{SG} < T < T_E$ (see Fig. 11). The thermal dependence of the exponent *α(T)* can be obtained considering that $\alpha(T) = -d(\ln(\chi))/d(\ln(T))$ (see inset Fig. 11).

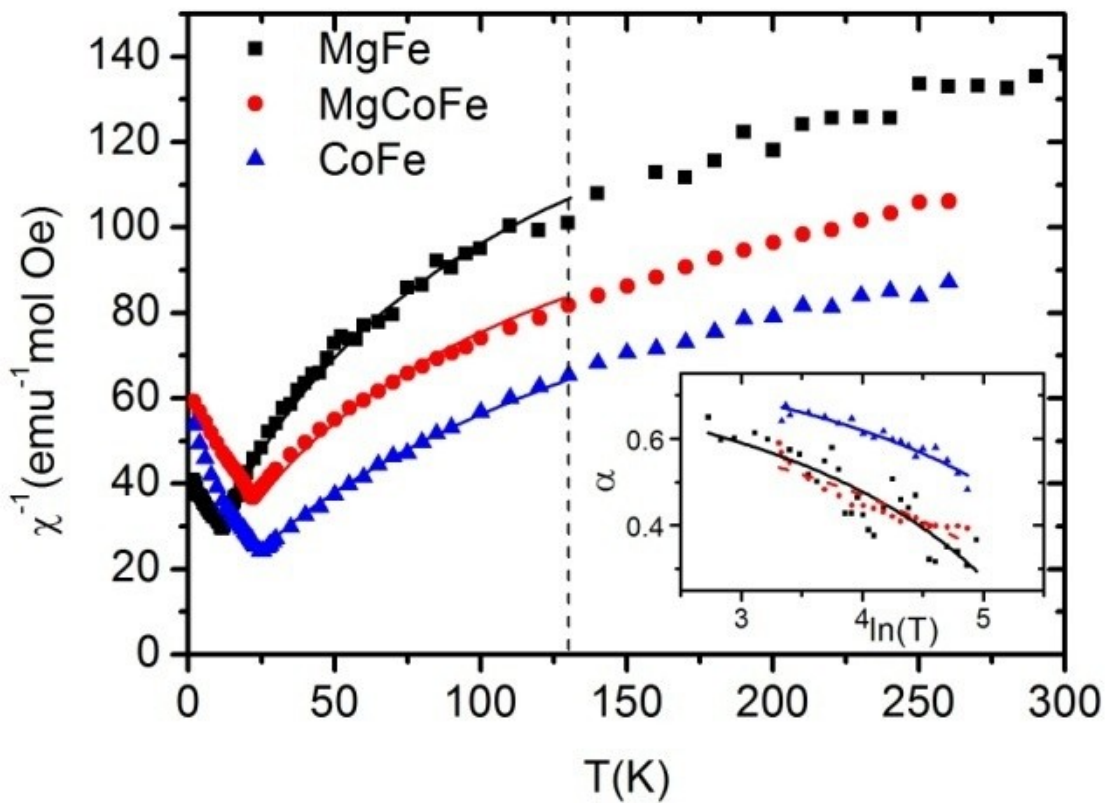

FIG. 11. (Color on line) Inverse of magnetic susceptibility versus temperature for $MgFeBO_4$ (black solid squares), $Mg_{0.5}Co_{0.5}FeBO_4$ (red solid circles) and $CoFeBO_4$ (blue solid triangles) showing the fit to a RSP (Eq. 4). Inset: Temperature dependence of the exponent *α(T)*. The line is the fit to the theoretical curve (see text).

The similar thermal dependence of the exponent *α(T),* with slowly varying functions are a signature that the three compounds are in a RSP in the intermediate temperature region[28]. Therefore, we can conclude that the susceptibility in the REHAC phase is characterized by a Random Singlet Phase behaviour.

### *4.4.3.High T: Paramagnetic regime*

Above $T_E$ there is no entanglement and the spin wave functions become factorizable and the magnetic susceptibility shows a paramagnetic Curie-Weiss behavior with a non-negligible temperature independent paramagnetism (TIP) contribution. This contribution can be attributed to a Van Vleck component of $Co^{2+}$ ions. From the fit of the $\chi^{-1}$ curve we can obtain the typical Curie-Weiss law parameters (see Table IV). In the fitting process typical $Co^{2+}$ TIP values, as obtained in the literature are considered [29]. Two sets of values of two different TIP values are shown in order to have an estimation of the variations of the fitted parameters.

The values obtained from the fit of the Mg-Fe warwickite are similar to those reported in the literature.[4] We observe an increasing trend in the *C* value as we increase the Co content, as should be expected for non interacting paramagnetic entities. The *θ* value is negative in all cases, and of the same order, indicating dominant antiferromagnetic interactions. The magnitude slightly increases when $Co^{2+}$ magnetic ions are present, although nothing can be asserted about the tendency in the three compounds, given the inaccuracy of the fitting procedure in this case.

TABLE IV. Curie-Weiss law fit parameters obtained from the $\chi^{-1}(T)$ in the high *T* regime. Estimated *C* value considering the spin states of the different ions is given for comparison (see text).

| | TIP (emu mol$^{-1}$) | *C* (emu K mol$^{-1}$) | *θ* (K) | Estimated *C* (emu K mol$^{-1}$) |
|---|---|---|---|---|
| MgFe | 0 | 4.0±0.4 | -283±30 | 4.37 |
| MgCoFe | 1 10$^{-4}$ | 5.3±0.4 | -317±30 | 5.31 |
| MgCoFe | 2 10$^{-4}$ | 5.1±0.4 | -302±30 | 5.31 |
| CoFe | 2 10$^{-4}$ | 6.5±0.4 | -315±30 | 6.25 |
| CoFe | 4 10$^{-4}$ | 6.2±0.4 | -307±30 | 6.25 |

It is important to estimate the expected values of the effective moment (expected *C* value) per formula unit in the paramagnetic phase for the studied set of warwickite compounds. We have considered that, the orbital component of magnetic moment is neglected and the spin component of the effective moment is calculated according with the formula: $\mu_S^2 = \sum_i g_i^2 S_i(S_i+1)$, accounting for the contribution of each type of transition ions. We assumed that all ions are in the high spin state and that all iron ions are in trivalent state. The spin values of magnetic ions are the following: ($Co^{2+}$: *S*=3/2, and $Fe^{3+}$: *S*=5/2), *g* = 2. There are one divalent ion and one trivalent ion per formula unit. Then for $MgFeBO_4$, $\mu_S = 5.916\,\mu_B$, giving an expected *C* value of 4.37. For $Mg_{0.5}Co_{0.5}FeBO_4$, $\mu_S = 6.52\,\mu_B$, $C$ = 5.31 emu K mol$^{-1}$ and for $CoFeBO_4$ $\mu_S = 7.07\,\mu_B$, $C$ = 6.25 emu K mol$^{-1}$.

## 5. SUPEREXCHANGE INTERACTION

To explain the magnetic behavior of warwickites under investigation, estimates of the superexchange interactions at *T*=0K are needed. We have used the simple model of superexchange interactions [30],[31] applied earlier to the analysis of the complex magnetic structure in $Co_3O_2BO_3$, $Co_2FeO_2BO_3$ ludwigites,[32],[33] and $Co_3B_2O_6$ cotoites [34], where it was found to describe the experimental results satisfactorily. The calculation is restricted by the nearest-neighbor approximation; i.e. only the interactions along the short M-O-M bonds are considered, while the long bonds M-O-M-O-M and M-O-B-O-M are neglected.

The warwickite structure has several types of indirect couplings: 93°, 95°, 98°, and 102°, which can be assigned to 90° exchange interactions, as well as 118° and 125° exchange interactions. They are described by nine exchange integrals *J*1-*J*9 (see Fig.

12). The $J1$-$J6$ are *intra-ribbon* interactions, while $J7$-$J9$ are *inter-ribbon* ones. In the 2-1-1-2 row the connected octahedra of the neighboring cations with common edges results in the exchange couplings with an angle of 98° ($J1$) and 95° ($J2$), respectively. The octahedra belonging to the adjacent rows, that are connected by a common edge, allow indirect couplings 98° ($J3$), 93-102° ($J4$, $J6$), and 95° ($J5$). The octahedra connected by a common oxygen ion belonging to the neighboring ribbons allow indirect couplings of 118° ($J7$, $J8$) and 125° ($J9$). The full set of the orbitals pairs participating in the coupling is listed in Table SMIII.

The antiferromagnetic (AF) and ferromagnetic (F) contributions from the six overlapping 3*d*-ligand-3*d* orbitals give rise to the superexchange integral *J*. The total integral of cation-cation exchange interaction *J* can be calculated as a sum of individual orbits exchange integrals

$$J = \frac{1}{4}\sum_{i,j=1}^{5(d)}\sum_{p=1}^{3}\frac{1}{S_i S_j} I_{ij}^{p} , \qquad (6)$$

where $S_{ij}$ - the interacting cations spins; the sum accounts for the five magnetic ion *d*-orbitals and three *p*-orbitals of the ligand; $I_{ij}^{p}$– the superexchange interaction integral between the individual orbitals *i, j* of two cations via oxygen p orbital. Interactions between two filled or two empty orbitals are neglected.

Taking into account superexchange bonds selected by lattice symmetry, one comes to the expressions for the exchange integrals corresponding to the cation pairs $Co^{2+}$-$Co^{2+}$, $Co^{2+}$-$Fe^{3+}$, $Fe^{3+}$-$Co^{2+}$, and $Fe^{3+}$-$Fe^{3+}$ (see Table SMIV). The calculated values of the cation-cation superexchange interaction are given in Table SMV.

In order to estimate the superexchange interactions in the studied warwickites we need to take into account the contributions of the different cations pairs $Co^{2+}$-$Co^{2+}$, $Co^{2+}$-$Fe^{3+}$, $Fe^{3+}$-$Co^{2+}$, $Fe^{3+}$-$Fe^{3+}$ to the total exchange integral. The site occupation factor as obtained from Mössbauer data [23] is used as a probability of each pair. We restrict this calculation to the MgFe and CoFe warwickites, as we do not have a precise cation distribution estimation for the MgFeCo warwickite. Detailed calculations are given in Supplementary Material.

### 5.1. $MgFeBO_4$

Both M1 and M2 ions are located in compressed oxygen octahedra. The singly occupied five *d*-orbitals of $Fe^{3+}$ ions interact antiferromagnetically. It leads to a negative value for all the $Fe^{3+}$-$Fe^{3+}$ integrals $J1$-$J9$ (see Table SMIII and Table SMV). The strongest interactions are intra-ribbon interactions $J1$-$J6$.

The crystallographic positions are divided into magnetic sublattices. The number of magnetic sublattices is determined by the different cations number, nonequivalent local cation positions number relative to the principal crystal axes, and interaction sign between the nearest neighbors at last. In the warwickites of interest the octahedra principal axes have four different directions relative to the cell axes. Let warwickite be considered as a magnetic system consisting of eight magnetic sublattices in which crystallographic positions M1 and M2 are divided into four magnetic sublattices: 1a, 1b, 1c, 1d and 2a, 2b, 2c, 2d (Fig. 12).

Calculated exchange interaction parameters in $MgFeBO_4$ are given in TableV. With these values, the mutual orientation of the sublattice magnetic moments are deduced and plotted in Figure 12.

TABLE V. The indirect exchange integrals (K) in the $MgFeBO_4$ and $CoFeBO_4$ warwickites.

| | Mg-Fe | Co-Fe |
|---|---|---|
| $J1$ | -1.26 | -3.82 |
| $J2$ | -1.89 | -3.26 |
| $J3$ | -1.26 | -3.82 |
| $J4$ | -1.89 | 0.15 |
| $J5$ | -1.89 | 0.15 |
| $J6$ | -0.84 | -2.50 |
| $J7$ | -0.42 | -2.26 |
| $J8$ | -0.42 | -1.96 |
| $J9$ | -0.52 | -2.83 |

The main results we found within the framework of our simple model calculation are that: i) in the $MgFeBO_4$ the strongest ordering antiferromagnetic interactions are the *intra-ribbon* ones coupling the cations along the *c*-axis ($J4$, $J6$) (see Fig. 12(a)). It leads to the appearance of the magnetic chains 2a-2c-2a, 1a-1c-1a, 1b-1d-1b, and 2b-2d-2b. ii) The net inter-chain interaction is negligible since the intensity of the ordering interactions $J3$, $J5$ and disordering ones $J1$, $J2$ are equal (see Table V). iii) There is doubling of the magnetic cell along the *c*-axis. It is necessary to note that a magnetic supercell with twice the volume of the structural cell was also found by neutron diffraction in $Mn_2BO_4$ warwickite.[7] iv) The *inter-ribbon* bond is strongly depressed due to frustrating interactions $J8$, $J9$ (Fig. 12(b)). The antiferromagnetic spin chains along *c*-axis and frustrating inter-chain bonds, as well as weak inter-ribbon interactions, do not allow the on-set of long range magnetic order.

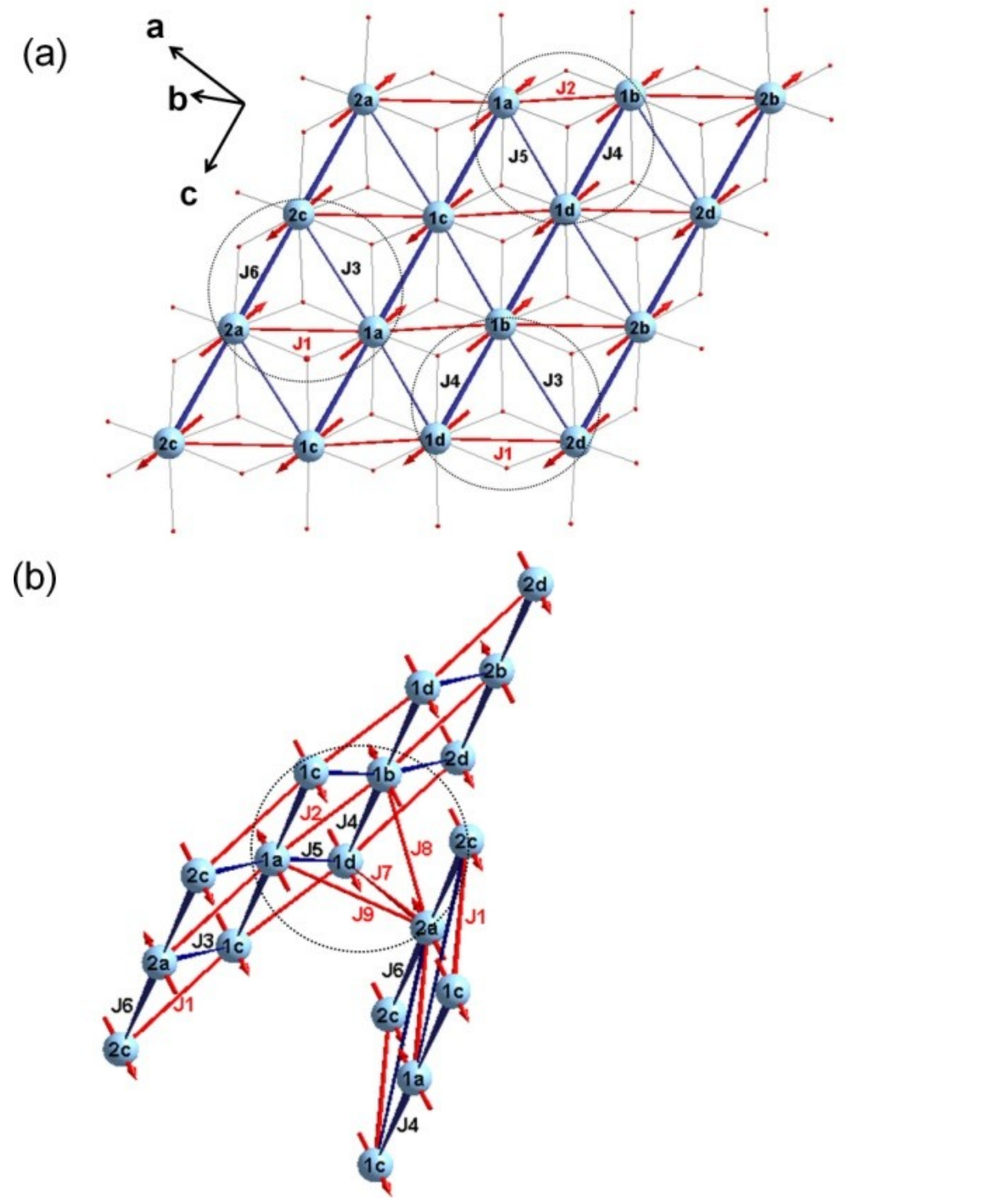

FIG. 12. (a) the *intra-ribbon* indirect exchange interactions ($J1$-$J6$) and b) *inter-ribbon* ones ($J7$-$J9$) in the $MgFeBO_4$ warwickite. Numerals indicate the belonging of a crystallographic position to a magnetic sublattice. The frustrated bonds are highlighted red. The interactions strength is shown by the lines thickness. The magnetic moments direction (randomly chosen relative to the crystallographic axes) demonstrate the ordering and disordering bonds. The non-equilateral triangles are highlighted by the circles.

### 5.2. *CoFeBO₄*

Let us consider the $Co^{2+}$-$Co^{2+}$ cation pair. For $Co^{2+}$ ions the $d_{xy}$ orbital is doubly occupied in a compressed octahedron. The seventh electron occupies with the same probability the $d_{xz}$, $d_{yz}$ orbitals, and each of these orbitals can be occupied either singly or doubly. The antiferromagnetic interactions $J1$, $J2$, $J3$ are considerably compensated by the ferromagnetic interactions induced by the overlapping of the singly occupied $d_{z^2}$, $d_{x^2-y^2}$ orbitals and doubly occupied $t_{2g}$ ones, as well as singly and doubly occupied $t_{2g}$ orbitals. The strongest interactions are those between the rows ($J4$, $J5$, $J6$) (see Table SMV). The orbitals overlap is such that all six contributions to the interaction have ferromagnetic nature, reinforcing the positive contribution to these integrals. The inter-ribbon interactions ($J7$, $J8$, $J9$) have predominantly antiferromagnetic character, which is enhanced by a negative contribution from the $e_g^1$ – O:2$p$ - $e_g^1$ orbitals overlap.

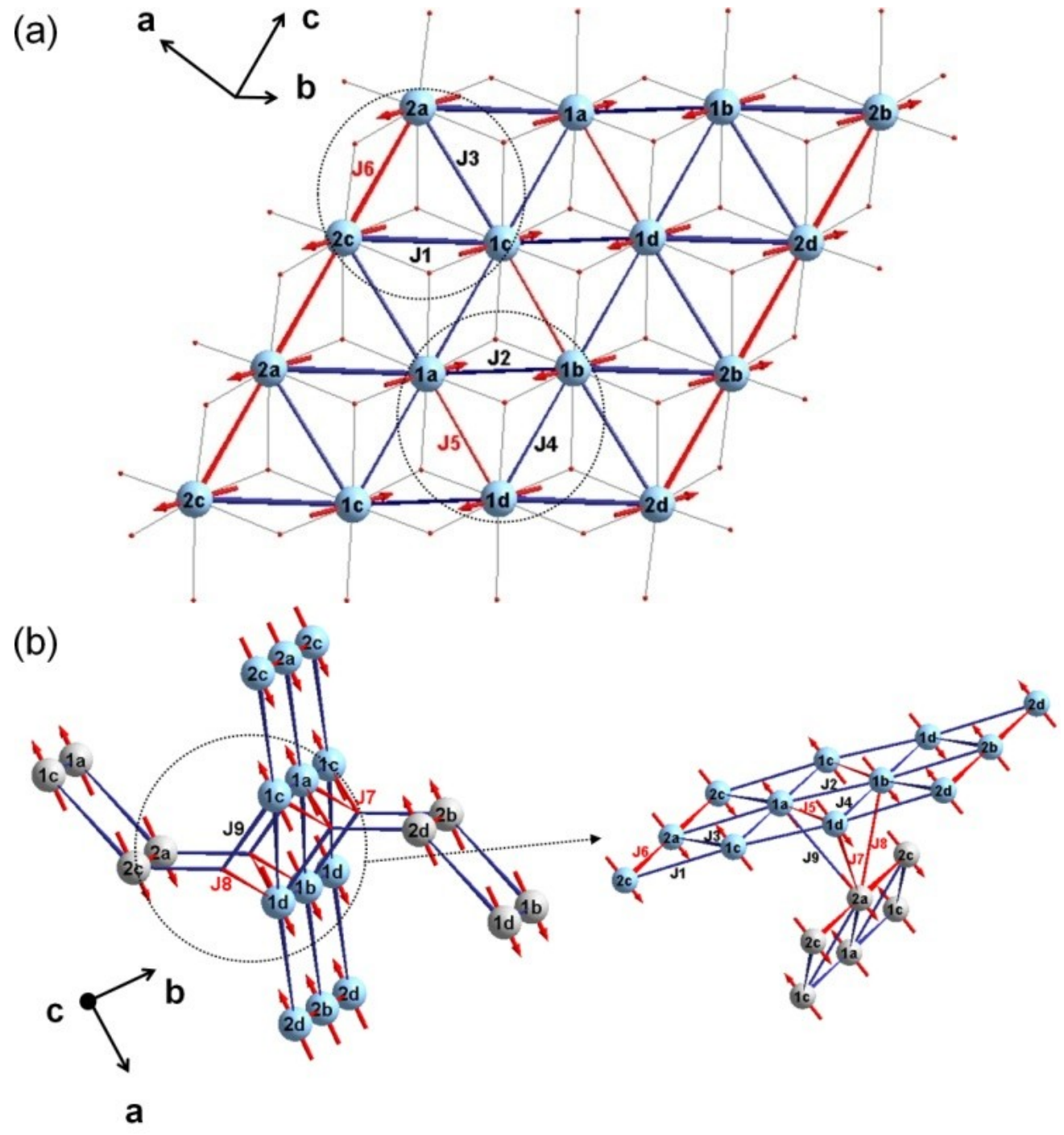

FIG. 13. (Color on line) The magnetic moments orientation obtained from the exchange interaction calculation and intra-ribbon indirect exchange (a) and the inter-ribbon ones (b) in the $CoFeBO_4$. The magnetic moment direction is arbitrarily chosen in *ac*-plane. The interactions strength is shown by the line's thickness. The frustrated bonds are highlighted red. The non-equilateral triangles are shown by the circles.

The calculated local magnetic structure, depicting the short range order, is presented in Fig. 13. The cations belonging to the magnetic sublattices 1a-1d are subject to the strong ordering exchange interaction from the adjacent sublattices 2a-2d. The negative interactions $J1$ and $J9$ reinforce each other and impose the magnetic structure (mutual orientation of magnetic moments). The antiferromagnetic interaction $J2$ and ferromagnetic one $J4$ support the AF structure inside the 1a-1d sublattices, while the $J5$ is a frustrating coupling (fig. 13(a)). The relatively strong disordering interactions $J6$ are active only within the sublattice 2a. The ordering interactions in the position M1 are stronger than the ones in the M2 position. At the same time, the strength of the disordering interactions in the M2 position is greater than that in the position M1.

In a molecular field approximation for the multisublattice model the exchange fields acting on the magnetic ions are defined by the competition between ordering and disordering interactions. For the $MgFeBO_4$ example, the estimations of the exchange fields $H_{exi}$ acting on the magnetic ions belonging to the $1a$ and $2a$ sublattices have given the values of $H_{ex}^{1a}$ = 60.7 and $H_{ex}^{2a}$ = 21.6 kOe, respectively. Such competition leads the magnetic moments at the different magnetic sites to become canted with respect to the average easy magnetization axis. The canting angle can change from site to site due to the variable molecular field. So, according to the simple superexchange interaction model the warwickites under investigation can be considered as non-collinear antiferromagnets where the canting angle of the magnetic moments has a random value.

## 6. DISCUSSION

In the following discussion, the main results on the physical properties studied in this work are summarized first, highlighting the spin-glass transition, observed magnetic anisotropy and the entanglement in the Random Singlet Phase. Then, the origin of magnetic anisotropy as due to the $Co^{2+}$ ion has been analyzed. The possible causes of the spin-glass state are presented and compared to other related compounds. Finally, the spin-glass state is interpreted in terms of the simple indirect coupling model of competing interactions.

The compounds studied in this work display a spin-glass transition at low temperatures, being $T_{SG}$ = 10 K for $MgFeBO_4$, $T_{SG}$ = 20 K for $Mg_{0.5}Co_{0.5}FeBO_4$ and $T_{SG}$ = 22 K for $CoFeBO_4$. There are many signs pointing to a spin-glass behavior: 1) the pronounced irreversibility in the FC/ZFC curves; 2) the flat low temperature dependence of the FC magnetization curve; 3) the non saturation of the magnetization even at magnetic fields as high as 140 kOe; 4) the observed thermo-remanence and the hysteresis loops shifted in magnetic field; 5) low temperature experiments have shown magnetic relaxation and memory effects in the thermo-remanence magnetization suggesting glassy dynamical properties associated with magnetic disorder and frustration. Besides, the analysis of the frequency dependence of the magnetic susceptibility cusp around $T_{SG}$ gives dynamical behavior parameters close to those of canonical spin-glasses. In addition, from temperature dependence Mössbauer experiments it has been found that at $T \leq T_{SG}$ the average hyperfine field fulfills $\langle H_{hf} \rangle_d \propto (T_{SG}-T)^{1/2}$, characteristic of short range spin-glasses [23].

Quantum entanglement appears at temperatures in between $T_{SG}$ and about $T_E$=130 K, confirming the existence of random magnetic chains, as in other heterometallic warwickite compounds [8]-[13]. At high temperatures ($T>T_E$), these systems follow a Curie-Weiss law with AF interactions. These AF couplings are due to intra-ribbon interactions giving rise to the low dimensional magnetic behavior at temperatures above the spin-glass transition. Moreover, it is slightly enhanced by the introduction of cobalt.

It is worth to underline the behaviour of these systems above $T_{SG}$ where random magnetic chains undergo fluctuations which are described under the Random Singlet Phase. The studied compounds are low-dimensional spin systems, perfect candidates to quantify quantum entanglement.

As stated in the Introduction, the main scope of this work is to analysis the influence of cobalt introduction in heterometallic warwickites. At this point, we can assert that the inclusion of magnetic Co ions in the series has the following effects: i) to increase of the spin-glass transition temperature, ii) to increase the magnetic net moment per formula unit, and iii) to induce an uniaxial anisotropy, which is neatly marked for the $CoFeBO_4$ system, where the *b*-axis is the easy axis of magnetization, while the *c*-axis the hardest magnetization axis. This anisotropy appears already in the paramagnetic state, increasing as cooling, and being maximal in the spin-glass regime. It can be attributed to a single ion anisotropy of the $Co^{2+}$ ion, which typically induces magnetic anisotropy due to the non-quenched orbital contribution of the ground state. Indeed, taking into account spin-orbit coupling, $Co^{2+}$ in a distorted octahedral field can be described by two Kramers doublets separated by about 100 $cm^{-1}$. At high temperatures the system behaves as an effective $S^*$=3/2 state with a residual orbital contribution which gives an effective momentum in between 4.7 and 5.2 $\mu_B$. At low temperatures, only the lowest Kramers doublet is populated. Orbital contribution from the nearest level results in a large anisotropy in the *g* value as the crystal field departs from cubic symmetry.

In Co-Fe warwickite, $Co^{2+}$ is in the center of an oxygen octahedra, similar to the coordination of cobalt ferrite [35]. The easy axis of magnetization of cobalt ferrite lies in the [100] direction, and its anisotropy is very large compared with other ferromagnetic ferrites, such as Mn, Fe and Ni ferrites, where the easy direction lies along the [111] axis. In general, the presence of $Co^{2+}$ ions in ferrites, induces a high anisotropy which always lies in the [100] direction. Moreover, the substitution of divalent metallic ions by a small amount of cobalt causes the change of easy direction of magnetization from [111] to [100]. So, we may expect by similarity to the ferrite case that this magnetic anisotropy arises from the low symmetry crystalline field of octahedral $Co^{2+}$ sites, due to the charge distribution caused by neighboring $Co^{2+}$ and $Fe^{3+}$ ions [35].

A striking feature of the magnetic properties of our compounds is a change in magnitude and anisotropy axis when substituting $Mg^{2+}$ ions partially or totally by $Co^{2+}$ ions. In Co-Fe warwickite, similarly to cobalt ferrite, charge distribution due to $Co^{2+}$ ions in the *ab* plane would induce the observed anisotropy with easy axis along *b* direction. In Mg-Co-Fe warwickite, however, with half $Co^{2+}$ ions, the probability to have a $Co^{2+}$ neighbor in the ribbon row is highly reduced, resulting in a reduction in the anisotropy of the magnetization, being the easy axis the *a* direction. Nevertheless, a small contribution is still observed, as evidenced by the secondary maximum observed in the $M(\theta)$ for the Mg-Co-Fe warwickite along the *b* axis (Fig. 4). In the structural study [24] it is found that Co addition gives rise to the distortion of $CoO_6$ octahedron, with M-O bond anisotropy increasing upon Co content. Therefore, induced anisotropy in Co warwickites may be associated to the modification of the $Co^{2+}$ crystal field due to the charge differences beyond the first coordination of oxygen atoms; i.e. because of the $Co^{2+}$ charges, as in ferrites.

Most hetero-metallic warwickites show typical spin-glass transition (Table I). All systems show high negative Weiss temperature $\theta$ and rather low magnetic ordering temperature $T_{SG}$. The former indicates the prevailing antiferromagnetic interactions. It has been proposed that the magnetic frustration level can be estimated using the ratio of $|\theta| / T_{SG}$ [36]. For instance, for ferromagnetic materials $|\theta| / T_c \sim 1$, for antiferromagnetic systems, $|\theta| / T_c \sim 2$-5. A high degree of frustration in a magnetic ordered system occurs for $|\theta| / T_c > 10$. For the majority of the warwickites of interest the value $|\theta|/ T_{SG}$ is in the range of 8 to 37 that are consistent with a high level of frustration. Interestingly, in $MnScBO_4$ both $\theta = -60$ K and $T_{SG} = 2.7$ K are much lower than for the $Mn_2BO_4$ and corresponding Fe - containing samples but the frustration ratio is still large 22.2. These values were found to be 28.3, 15.5 and 14 for $MgFeBO_4$, $Mg_{0.5}Co_{0.5}FeBO_4$ and $CoFeBO_4$ respectively. Therefore a high degree of frustration is present in the studied warwickites.

Using a simple indirect coupling model, disregarding other exchange mechanisms and the magnetic anisotropy, we have calculated the exchange integrals in two Mg, Co, and Fe- containing warwickites and offered a simple scenario of the magnetic interactions. According to this scheme strong intra-ribbon exchange is dominant giving rise to the low dimensional phase, which can be classified as a RSP. The weaker inter-ribbon couplings and a high level of magnetic frustration set on spin-glass behavior below $T_{SG}$.

Indeed, it is well known that the spin-glass behavior is a result of the randomness of the value and sign of the exchange interactions and can be caused by crystallographic or magnetic disorder, and frustration. The latter is found when competing interactions between the magnetic moments in a triangular lattice are effective. Previous structural analysis [24] clearly indicates the existence of atomic disorder in all three warwickites under investigation. In addition, several types of triangular motifs can be distinguished both inside the ribbon and between the adjacent ribbons (see Fig. SMI and Fig. 12). Three isosceles triangles are resolved inside of the ribbon involving different exchange couplings $J1$-$J3$-$J6$, $J2$-$J4$-$J5$ and $J1$-$J3$-$J4$ (Fig. 12(a)). A bit more complex bond geometry exists between the adjacent ribbons. Three types of triangles can be singled out: one is the isosceles triangle $J4$-$J7$-$J8$ and the other two are scalene triangles with exchange couplings $J2$-$J8$-$J9$ and $J5$-$J7$-$J9$. At least one out of three exchange bonds in the triangles, both inside the ribbon and between them, induces frustration. The mutual orientation of the magnetic moments predicted with the calculated exchange AF integrals $J2$, $J8$, and $J9$ inside the non-equilateral triangle help to create frustration. All this indicates high level of frustration in the Fe-containing warwickite.

In $MgFeBO_4$ there is just one type of magnetic ion $Fe^{3+}$. If all metallic sites were occupied by $Fe^{3+}$ ions, the magnetic frustration level would be high since the ordering and disordering AF bonds are almost equal in number (see Table SMV). The Mg addition breaks the magnetic bonds and leads to a decrease in magnetic frustration degree. Experimentally it is expressed as spin-glass behavior with relatively low $T_{SG}$ =10 K. The strongest ordering antiferromagnetic interactions $J4$, $J6$ give rise to the doubling of the magnetic cell along the $c$-axis. The magnetic structure of $MgFeBO_4$ can be represented by antiferromagnetic $Fe^{3+}$ chains extended along the $c$-axis. The magnetic coupling between the adjacent chains is weakened due to disordering interactions $J1$, $J2$, $J8$, $J9$. This feature leads to the effective magnetic quasi $1D$ structure of $MgFeBO_4$. The antiferromagnetic spin chains along $c$-axis and frustrating inter-chain bonds, as well as weak inter-ribbon interactions, favor the spin-glass state.

When $Co^{2+}$ ($S$ = 3/2) substitutes for diamagnetic $Mg^{2+}$, $T_{SG}$ increases up to 22 K. Though the inter-ion distances in the triangles remain almost unchanged [24] the $Co^{2+}$ addition changes the coupling signs, and brings about a change of the exchange integrals values. The magnitude of the exchange interactions ($J$) increases (Table V). The substitution of $Fe^{3+}$ ($S$ = 5/2) by diamagnetic $Mg^{2+}$ decreases considerably the average spin $\langle S \rangle$ per site in $MgFeBO_4$. On the contrary, in $CoFeBO_4$ both positions are occupied by magnetic ions $Fe^{3+}$ and $Co^{2+}$ ($S$ = 3/2), that induces an increase in the average spin. Both the exchange integral $J$ and spin $\langle S \rangle$ actually determine the exchange energy and can give rise to an increment in $T_{SG}$ in $CoFeBO_4$. The level of spin frustration in $CoFeBO_4$ remains high ($|\theta| / T_{SG} \approx$ 14) but is smaller than that in $MgFeBO_4$. A rough estimation of the ratio of frustrating to the total number of exchange couplings is ~ 40 % in $MgFeBO_4$ and only ~30 % in $CoFeBO_4$. One can see that inside the ribbon two kinds of triangles with one frustrating bond are formed (Fig. 12(a)). Along with AF interactions, the FM ones $J4$ and $J5$ exist. The strong AF interactions ($J1$, $J2$ and $J3$) and FM interaction $J4$ gives rise to AF ordering coupling in the row 2-1-1-2, with ferromagnetic coupling between the rows. The interactions between the adjacent ribbons $J7$, $J8$, $J9$ are weaker than those inside the ribbons $J1$-$J6$ due to the fact that the pathways consist of common oxygen atom and M-O-M angles 118° and 125°. In the triangles connecting adjacent ribbons with the bonds of $J4$-$J7$-$J8$ and $J5$-$J7$-$J9$, the exchange interactions are doubly frustrating. On the other hand, the triangle with the bonds $J2$-$J8$-$J9$ has just one frustrating interaction (Fig. 12(b)).

## 7. CONCLUSIONS

The warwickite structure of $MgFeBO_4$, $Mg_{0.5}Co_{0.5}FeBO_4$ and $CoFeBO_4$ warwickites is formed by weakly coupled magnetic ribbons. They display a spin-glass behavior at low temperatures, showing magnetic anisotropy in the Co substituted compounds. The three compounds show quantum entanglement behavior $\chi(T) \alpha \; T^{-\alpha}$ between $T_{SG}$ the spin-glass transition temperature, and $T_E$, the entanglement temperature region threshold, (intermediate region). The α parameters have been deduced as a function of temperature and Co, indicating the existence of random singlet phase in this temperature region.

Our results points to the randomness in the crystal site occupation; i.e. intrinsic disorder due to the presence of different metal ions and disordered substitutional atomic arrangement, and the presence of triangular motifs with competing interactions due to the crystal structure of the warwickite as the main causes for the low temperature spin-glass behavior of these systems. Indeed, the strong competing AF interactions among the magnetic moments in the triangles leads to high frustration level and does not allow the on-set of long magnetic order.

We may conclude that these compounds undergo a spin-glass transition that is caused by spin short range correlations, with frustration and chemical disorder as the mechanisms governing the transition.

The introduction of $Co^{2+}$ induces uniaxial anisotropy since a preferred magnetization direction is imposed by the crystalline field. The different magnetic easy axis directions in Mg-Co-Fe and Co-Fe compounds is attributed to different charge distribution of neighboring $Co^{2+}$ ions. The substitution of $Mg^{2+}$ by $Co^{2+}$ has the additional effect of increasing the net exchange interaction, resulting in a higher spin-glass transition temperature and a lower degree of frustration.

**ACKNOWLEDGMENTS**

This work has been financed by the MECOM Project MAT11/23791, MAT2014-53921-R and DGA IMANA project E-34, Russian Foundation for Basic Research(projects №12-02-31543-mol-a, 13-02-00958, 13-02-00358-a), Council for Grants of the President of the Russian Federation (project nos. NSh-2886.2014.2, SP-938.2015.5). The work of one of coauthors (M.S.P.) was supported by the Krasnoyarsk Regional Fund of Science and Technical Activity Support, and by the program of Foundation for Promotion of Small Enterprises in Science and Technology ("UMNIK" program).

*Corresponding author: aarauzo@unizar.es

# Supplementary Material for: Spin-glass behavior in single crystals of heterometallic magnetic warwickites $MgFeBO_4$, $Mg_{0.5}Co_{0.5}FeBO_4$, and $CoFeBO_4$

A. Arauzo[1*], N.V. Kazak[2], N.B. Ivanova[3], M.S. Platunov[2], Yu.V. Knyazev[3], O.A. Bayukov[2], L.N. Bezmaternykh[2], I.S. Lyubutin[4], K.V. Frolov[4], S.G. Ovchinnikov[2,3,5], and J. Bartolomé[6]

[1]*Servicio de Medidas Físicas. Universidad de Zaragoza, Pedro Cerbuna 12, 50009 Zaragoza, Spain.*

[2]*L.V. Kirensky Institute of Physics, SB of RAS, 660036 Krasnoyarsk, Russia*

[3]*Siberian Federal University, 660074 Krasnoyarsk, Russia*

[4]*Shubnikov Institute of Crystallography, RAS, 119333, Moscow, Russia*

[5]*Siberian State Aerospace University, 660014 Krasnoyarsk, Russia*

[6]*Instituto de Ciencia de Materiales de Aragón. CSIC-Universidad de Zaragoza and Departamento de Física de la Materia Condensada. 50009 Zaragoza, Spain*

## SM1. Structural information

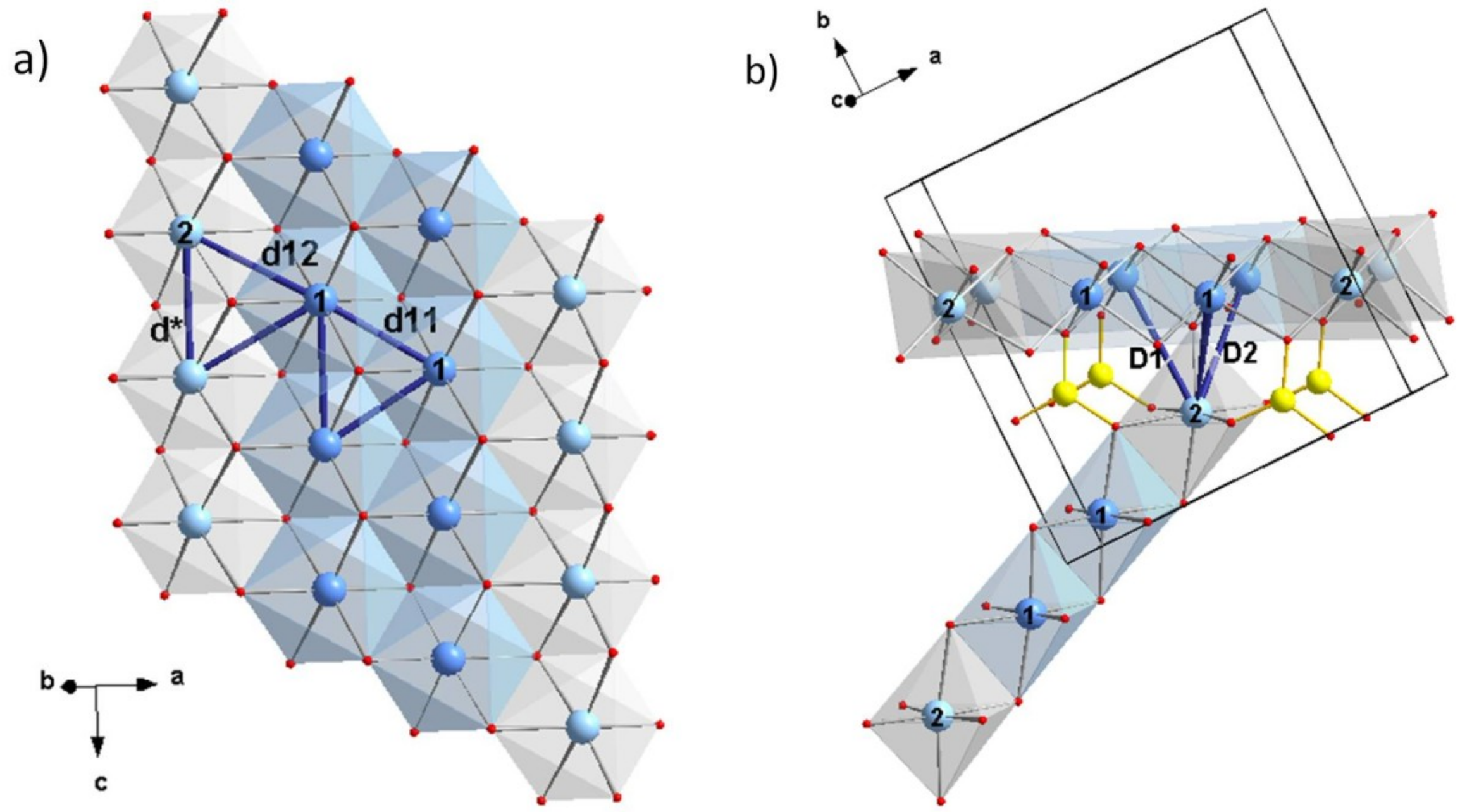

FIG. SMI (a) The rows formed by four edge sharing octahedra stacked in the sequence 2 – 1 – 1 - 2 are located across the structural subunits (ribbons). The intra-ribbon distances between metal ions are shown, (b) the closest four-octahedra flat ribbons. The trigonal group ($BO_3$) located in the voids between the ribbons and the shortest inter-ribbon distance between metal ions are displayed. Metal distances are given in Table SMII.

TABLE SMI. Crystal data for Mg-Fe, Mg-Co-Fe and Co-Fe warwickites at 296 K

| | Mg-Fe | Mg-Co-Fe | Co-Fe |
|---|---|---|---|
| Formula weight (g mol$^{-1}$) | 154.97 | 172.27 | 189.59 |
| Crystal system | | orthorhombic | |
| Space-group | | *Pnma* (62) | |
| Unit cell parameters (Å) | $a$ = 9.2795(10)<br>$b$ = 9.4225(10)<br>$c$ = 3.1146(3) | $a$ = 9.2449<br>$b$ = 9.3898<br>$c$ = 3.1185 | $a$ = 9.2144<br>$b$ = 9.3651<br>$c$ = 3.1202 |
| Unit cell volume (Å$^3$) | 272.33(5) | 270.71(6) | 269.25 |
| Z | | 2 | |
| Calculated density (g cm$^3$) | 1.89 | 2.11 | 2.34 |
| Crystal shape | | Needle (along $c$) | |

TABLE SMII. The closest distances (Å) between metal ions in the warwickite structure [1].

| | $d_{11}$ | $d_{12}$ | $d^*$ | $D_1$ | $D_2$ |
|---|---|---|---|---|---|
| $MgFeBO_4$ | 2.9511 | 3.2592 | 3.1146 | 3.5646 | 3.3996 |
| $Mg_{0.5}Co_{0.5}FeBO_4$ | 2.9554 | 3.2419 | 3.1185 | 3.5347 | 3.3978 |
| $CoFeBO_4$ | 2.9529 | 3.2296 | 3.1202 | 3.5135 | 3.3929 |

## SM2. Calculation of Superexchange Interactions

TABLE SMIII. The orbitals pairs participating in the superexchange interactions in Mg-Fe, Mg-Co-Fe, Co-Fe warwickites. The arrows denote the direction of electron transfer between interacted orbitals. The superscripts mean the indirect bond angles; the subscripts mean the crystallographic position numbers.

| Exchange integral | Orbitals pair | Ions type | | | | Belonging |
|---|---|---|---|---|---|---|
| | | **$Co^{2+}$-$Co^{2+}$** | **$Co^{2+}$-$Fe^{3+}$** | **$Fe^{3+}$-$Co^{2+}$** | **$Fe^{3+}$-$Fe^{3+}$** | |
| $J1 = J_{12}^{98°}$ | $dyz \to dz^2$ | AF | AF | AF | AF | Row |
| | $dz^2 \to dxz$ | F | AF | F | AF | |
| | $dxz \to dxy$ | F | F | F | AF | |
| | $dxy \to dyz$ | F | F | AF | AF | |
| | $dxz \to dx^2\text{-}y^2$ | F | F | AF | AF | |
| | $dx^2\text{-}y^2 \to dyz$ | AF | AF | AF | AF | |
| $J2 = J_{11}^{95°}$ | $dz^2 \to dxz$ | F | AF | F | AF | Row |
| | $dyz \to dz^2$ | AF | AF | AF | AF | |
| | $dxz \to dxy$ | F | F | F | AF | |
| | $dxy \to dyz$ | F | F | AF | AF | |
| | $dxz \to dx^2\text{-}y^2$ | F | F | AF | AF | |
| | $dx^2\text{-}y^2 \to dyz$ | AF | AF | AF | AF | |
| $J3 = J_{12}^{98°}$ | $dyz \to dz^2$ | AF | AF | AF | AF | Ribbon |
| | $dz^2 \to dxz$ | F | AF | F | AF | |
| | $dxz \to dxy$ | F | F | F | AF | |
| | $dxy \to dyz$ | F | F | AF | AF | |
| | $dx^2\text{-}y^2 \to dxz$ | F | F | F | AF | |
| | $dyz \to dx^2\text{-}y^2$ | AF | AF | AF | AF | |
| $J4 = J_{11}^{93,102°}$ | $dxy \to dx^2\text{-}y^2$ | F | F | AF | AF | Ribbon |
| | $dx^2\text{-}y^2 \to dxy$ | F | AF | F | AF | |
| | $dxy \to dz^2$ | F | F | AF | AF | |
| | $dz^2 \to dxy$ | F | AF | F | AF | |
| | $dyz \to dxz$ | F | AF | F | AF | |
| | $dxz \to dyz$ | F | F | AF | AF | |
| $J5 = J_{11}^{95°}$ | $dxy \to dx^2\text{-}y^2$ | F | F | AF | AF | Ribbon |
| | $dx^2\text{-}y^2 \to dxy$ | F | AF | F | AF | |
| | $dxy \to dz^2$ | F | F | AF | AF | |
| | $dz^2 \to dxy$ | F | AF | F | AF | |
| | $dyz \to dxz$ | F | AF | F | AF | |
| | $dxz \to dyz$ | F | F | AF | AF | |
| $J6 = J_{22}^{93,102°}$ | $dx^2\text{-}y^2 \leftrightarrow dxy$ | F | AF | F | AF | Ribbon |
| | $dz^2 \leftrightarrow dxy$ | F | AF | F | AF | |
| | $dyz \leftrightarrow dxz$ | F | AF | F | AF | |
| $J7 = J_{12}^{118°}$ | $dz^2 \leftrightarrow dz^2$ | AF | AF | AF | AF | Interribbon |
| | $dxz \to dxz$ | F | F | F | AF | |
| | $dyz \leftrightarrow dyz$ | AF | AF | AF | AF | |
| | $dx^2\text{-y}^2 \to dz^2$ | AF | AF | AF | AF | |
| $J8 = J_{12}^{118°}$ | $dz^2 \leftrightarrow dz^2$ | AF | AF | AF | AF | Interribbon |
| | $dz^2 \to dx^2\text{-y}^2$ | AF | AF | AF | AF | |
| | $dxz \leftrightarrow dxz$ | F | F | F | AF | |
| | $dxy \to dyz$ | F | F | AF | AF | |
| $J9 = J_{12}^{125°}$ | $dz^2 \leftrightarrow dz^2$ | AF | AF | AF | AF | Interribbon |
| | $dxz \to dxz$ | F | F | F | AF | |
| | $dyz \leftrightarrow dyz$ | AF | AF | AF | AF | |
| | $dx^2\text{-}y^2 \to dz^2$ | AF | AF | AF | AF | |

TABLE SMIV. The expressions for the superexchange integrals inside the cation pairs. Here, *b* and *c* are the parameters of electron transfer along the $\sigma$ and $\pi$ bonds, respectively; *U* – is the energy of ligand-cation excitation; and *J* is the intra-atomic exchange integral (Hund energy). The parameters values are *b* = 0.02, *c* = 0.01 and $U_{Co^{3+}}$ = 3.2 eV, $U_{Fe^{3+}}$ = 4.5 eV, $J_{Co^{2+}}$ = 3.0 eV for a spinel structure in which the octahedral positions have inter-ionic distances comparable with those in warwickite [2],[3]. The factor $|\cos(\theta)|$ takes into account the indirect bond angles. The sin($\theta$) allows unambiguously to determine the mutual arrangement of the iron magnetic moment in the structure.

**$Co^{2+} – Co^{2+}$**

$$J1 = J3 = \frac{1}{9}\left[\frac{8}{3}bc\left(J_{Co^{2+}} - 2U_{Co^{2+}}\right) + 2c^2 J_{Co^{2+}}\right]$$

$$J2 = \frac{1}{9}\left[\frac{8}{3}bc\left(J_{Co^{2+}} - 2U_{Co^{2+}}\right) + 2c^2 J_{Co^{2+}}\right]$$

$$J4 = \frac{1}{9}c\left[\left(\frac{16}{3}b + 2c\right)J_{Co^{2+}}\right]$$

$$J5 = \frac{1}{9}c\left[\left(\frac{16}{3}b + 2c\right)J_{Co^{2+}}\right]$$

$$J6 = \frac{1}{9}c\left[\frac{16}{3}b + 2c\right]J_{Co^{2+}}$$

$$J7 = -\frac{1}{9}\left[\left(\frac{16}{9}b^2 + c^2\right)2U_{Co^{2+}} - c^2 J_{Co^{2+}}\right]\left|\cos(118°)\right|$$

$$J8 = -\frac{1}{9}\left[\frac{16}{9}b^2 2U_{Co^{2+}} - 2c^2 J_{Co^{2+}}\right]\left|\cos(118°)\right|$$

$$J9 = -\frac{1}{9}\left[\left(\frac{16}{9}b^2 + c^2\right)2U_{Co^{2+}} - c^2 J_{Co^{2+}}\right]\left|\cos(125°)\right|$$

**$Co^{2+} – Fe^{3+}$**

$$J1 = J3 = -\frac{1}{15}c\left[\frac{13}{3}b\left(U_{Co^{2+}} + U_{Fe^{3+}}\right) - (b + 2c)J_{Co^{2+}}\right]$$

$$J2 = -\frac{1}{15}c\left[\frac{13}{3}b\left(U_{Co^{2+}} + U_{Fe^{3+}}\right) - (b + 2c)J_{Co^{2+}}\right]$$

$$J4 = -\frac{1}{15}\left[\left(\frac{8}{3}bc + c^2\right)\left(U_{Co^{2+}} + U_{Fe^{3+}} - J_{Co^{2+}}\right)\right]$$

$$J5=-\frac{1}{15}\left[\left(\frac{8}{3}bc+c^2\right)\left(U_{Co^{2+}}+U_{Fe^{3+}}-J_{Co^{2+}}\right)\right]$$

$$J6=-\frac{1}{15}\left[\left(\frac{8}{3}bc+c^2\right)\left(U_{Co^{2+}}+U_{Fe^{3+}}-J_{Co^{2+}}\right)\right]$$

$$J7=-\frac{1}{15}\left[\left(\frac{16}{9}b^2+c^2\right)\left(U_{Co^{2+}}+U_{Fe^{3+}}\right)-c^2J_{Co^{2+}}\right]\left|\cos(118°)\right|$$

$$J8=-\frac{1}{15}\left[\frac{16}{9}b^2\left(U_{Co^{2+}}+U_{Fe^{3+}}\right)-2c^2J_{Co^{2+}}\right]\left|\cos(118°)\right|$$

$$J9=-\frac{1}{15}\left[\left(\frac{16}{9}b^2+c^2\right)\left(U_{Co^{2+}}+U_{Fe^{3+}}\right)-c^2J_{Co^{2+}}\right]\left|\cos(125°)\right|$$

**$Fe^{3+}$ – $Co^{2+}$**

$$J1=-\frac{1}{15}c\left[\left(\frac{11}{3}b+c\right)\left(U_{Co^{2+}}+U_{Fe^{3+}}\right)-\left(\frac{5}{3}b+c\right)J_{Co^{2+}}\right]$$

$$J2=-\frac{1}{15}c\left[\left(\frac{11}{3}b+c\right)\left(U_{Co^{2+}}+U_{Fe^{3+}}\right)-\left(\frac{5}{3}b+c\right)J_{Co^{2+}}\right]$$

$$J3=-\frac{1}{15}c\left[\left(\frac{11}{3}b+c\right)\left(U_{Co^{2+}}+U_{Fe^{3+}}\right)-\left(\frac{5}{3}b+c\right)J_{Co^{2+}}\right]$$

$$J4=-\frac{1}{15}c\left[\left(\frac{8}{3}b+c\right)\left(U_{Co^{2+}}+U_{Fe^{3+}}-J_{Co^{2+}}\right)\right]$$

$$J5=-\frac{1}{15}c\left[\left(\frac{8}{3}b+c\right)\left(U_{Co^{2+}}+U_{Fe^{3+}}-J_{Co^{2+}}\right)\right]$$

$$J6=-\frac{1}{15}c\left[\left(\frac{8}{3}b+c\right)\left(U_{Co^{2+}}+U_{Fe^{3+}}-J_{Co^{2+}}\right)\right]$$

$$J7=-\frac{1}{15}\left[\left(\frac{16}{9}b^2+c^2\right)\left(U_{Co^{2+}}+U_{Fe^{3+}}\right)-c^2J_{Co^{2+}}\right]\left|\cos(118°)\right|$$

$$J8=-\frac{1}{15}\left[\frac{16}{9}b^2\left(U_{Co^{2+}}+U_{Fe^{3+}}\right)-2c^2J_{Co^{2+}}\right]\left|\cos(118°)\right|$$

$$J9=-\frac{1}{15}\left[\left(\frac{16}{9}b^2+c^2\right)\left(U_{Co^{2+}}+U_{Fe^{3+}}\right)-c^2J_{Co^{2+}}\right]\left|\cos(125°)\right|$$

**$Fe^{3+}$ – $Fe^{3+}$**

$$J1=J3=-\frac{4}{75}c\left[(8b+c)U_{Fe^{3+}}\right]\sin(98°)$$

$$J2=J5=-\frac{4}{75}c\left[(8b+3c)\left(U_{Fe^{3+}}\right)\right]\sin(95°)$$

$$J4=-\frac{4}{75}c\left[(8b+3c)\left(U_{Fe^{3+}}\right)\right]\sin(97°)$$

$$J6=-\frac{4}{75}c\left[(8b+3c)\left(U_{Fe^{3+}}\right)\right]\sin(97°)$$

$$J7=J8=-\frac{4}{25}\left[\left(\frac{8}{9}b^2+c^2\right)\left(U_{Fe^{3+}}\right)\right]\left|\cos(118°)\right|$$

$$J9 = -\frac{4}{25}\left[\left(\frac{8}{9}b^2 + c^2\right)\left(U_{Fe^{3+}}\right)\right]\left|\cos(125°)\right|$$

TABLE SMV. Superexchange integrals (K) of the cation-cation interactions in Mg-Fe, Mg-Co-Fe, and Co-Fe warwickites.

| | $Co^{2+}$-$Co^{2+}$ | $Co^{2+}$-$Fe^{3+}$ | $Fe^{3+}$-$Co^{2+}$ | $Fe^{3+}$-$Fe^{3+}$ |
|---|---|---|---|---|
| *J*1 | -1.56 | -4.23 | -3.23 | -5.23 |
| *J*2 | -1.56 | -4.23 | -3.23 | -5.27 |
| *J*3 | -1.56 | -4.23 | -3.23 | -5.23 |
| *J*4 | 4.89 | -2.3 | -2.3 | -5.25 |
| *J*5 | 4.89 | -2.3 | -2.3 | -5.27 |
| *J*6 | 4.89 | -2.3 | -2.3 | -5.25 |
| *J*7 | -2.96 | -2.15 | -2.15 | -1.73 |
| *J*8 | -2.39 | -1.77 | -1.77 | -1.73 |
| *J*9 | -3.61 | -2.63 | -2.63 | -2.18 |

*$MgFeBO_4$*

The exchange interaction parameters in $MgFeBO_4$ (Table V) are calculated using the numbers of nearest neighbors $z_{ij}$ and site occupation factor $x_{Fe}$ obtained from the Mössbauer results, by means of the expression $Ji = \sum_{j=n.n} z_{ij} x_{Fei} x_{Fej} Ji_{Fe^{3+}-Fe^{3+}}$, where $Ji_{Fe^{3+}-Fe^{3+}}$ for $i$ = 1, 9, are values taken from Table SMV. The mutual orientation of the sublattice magnetic moments are deduced using the calculated exchange integrals in $MgFeBO_4$. The relative orientation of the different sublattice moments are shown in the Table SMVI by the arrows. The exchange integrals of minimum energy correspond to the strongest coupling. For example, for the atom 1a↑ (first column, first row), the strongest interaction is AF with the atom 1c↓ n.n. Fe (first column, third row), via two *J*4 exchange paths (-3.78 K). Therefore, this interaction tends to order the magnetic moments of the 1a and 1c sublattices antiferromagnetically (we name this type of coupling as “ordering interaction”, is denoted in Table SMVI in bold, and in Fig. 13 with black lines). These interactions establish the directions of the arrows. In contrast, first column 1a↑ is AF coupled to second row 1b↑, so it is opposing the ↑↑ predominant ordering arrangement, thus “frustrating” the ordering (these couplings are named as “disordering interaction”, is denoted in Table SMVI in italic, and in Fig. 13 with red lines).

TABLE SMVI. The exchange interactions integrals (K) in $MgFeBO_4$ warwickite. The strongest ordering interactions are shown in bold. The disordering (frustrating) interactions are shown in italic.

| | 1a ↑ | 1b ↑ | 1c ↓ | 1d ↓ | 2a ↑ | 2b ↑ | 2c ↓ | 2d ↓ |
|---|---|---|---|---|---|---|---|---|
| 1a ↑ | 0 | *-1.89* | **-3.78** | -1.89 | *-1.78* | *-0.42* | -1.26 | -0.42 |
| 1b ↑ | *-1.89* | 0 | -1.89 | **-3.78** | *-0.42* | *-1.78* | -0.42 | -1.26 |
| 1c ↓ | **-3.78** | -1.89 | 0 | *-1.89* | -1.26 | -0.42 | *-1.78* | *-0.42* |
| 1d ↓ | -1.89 | **-3.78** | *-1.89* | 0 | -0.42 | -1.26 | *-0.42* | *-1.78* |
| 2a ↑ | *-1.78* | *-0.42* | -1.26 | -0.42 | 0 | | **-1.68** | |
| 2b ↑ | *-0.42* | *-1.78* | -0.42 | -1.26 | | 0 | | **-1.68** |
| 2c ↓ | -1.26 | -0.42 | *-1.78* | *-0.42* | **-1.68** | | 0 | |
| 2d ↓ | -0.42 | -1.26 | *-0.42* | *-1.78* | | **-1.68** | | 0 |

*$CoFeBO_4$*

To estimate the superexchange interactions in the Co-containing warwickites we need to take into account the contributions of the different cations pairs $Co^{2+}$-$Co^{2+}$, $Co^{2+}$-$Fe^{3+}$, $Fe^{3+}$-$Co^{2+}$, $Fe^{3+}$-$Fe^{3+}$ to the total exchange integral. The site occupation factor defined from the Mössbauer data was used as a probability of each pair. For example, the exchange interactions integral $J1$ was calculated by applying the simple expression

$$J1 = x_{Co1}x_{Co2}J1_{Co^{2+}-Co^{2+}} + x_{Co1}x_{Fe2}J1_{Co^{2+}-Fe^{3+}} + x_{Fe1}x_{Co2}J1_{Fe^{3+}-Co^{2+}} + x_{Fe1}x_{Fe2}J1_{Fe^{3+}-Fe^{3+}} \quad (7)$$

where $x_{M1}$, $x_{M2}$ are the concentrations of Co and Fe ions in the M1 and M2 crystallographic positions. The exchange interactions obtained for Co-Fe warwickite are listed in Tables V and SMVII. The magnetic moment directions predicted from the calculated exchange parameters are shown by the arrows.

TABLE SMVII. The exchange interaction integrals (K) in the $CoFeBO_4$ warwickite. The strongest ordering interactions are shown in bold. The disordering (frustrating) interactions are shown in italic.

| | 1a↑ | 1b↓ | 1c↑ | 1d↓ | 2a↓ | 2b↑ | 2c↓ | 2d↑ |
|---|---|---|---|---|---|---|---|---|
| 1a↑ | 0 | -3.26 | +0.3 | *+0.15* | **-6.65** | *-1.96* | -3.82 | *-2.26* |
| 1b↓ | -3.26 | 0 | *+0.15* | +0.3 | *-1.96* | **-6.65** | *-2.26* | -3.82 |
| 1c↑ | +0.3 | *+0.15* | 0 | -3.26 | -3.82 | *-2.26* | **-6.65** | -1.96 |
| 1d↓ | *+0.15* | +0.3 | -3.26 | 0 | *-2.26* | -3.82 | *-1.96* | **-6.65** |
| 2a↓ | **-6.65** | *-1.96* | -3.82 | *-2.26* | *0* | | *-5.0* | |
| 2b↑ | *-1.96* | **-6.65** | *-2.26* | -3.82 | | *0* | | *-5.0* |
| 2c↓ | -3.82 | *-2.26* | **-6.65** | *-1.96* | *-5.0* | | *0* | |
| 2d↑ | *-2.26* | -3.82 | *-1.96* | **-6.65** | | *-5.0* | | *0* |

## SM3. Heat Capacity Measurements

The lattice phonon contribution has been removed to obtain the magnetic contribution which is shown in Fig. SM2. In the $MgFeBO_4$ compound, a broad maximum at about $T = 1.3$ $T_{SG}$ is clearly observed, while the magnetic contribution of the Mg-Fe-Co compound spreads over a higher temperature range, resulting in a much less pronounced maximum. There is no abrupt long-range order anomalous peak, as corresponds to a spin-glass phase transition. The application of a magnetic field has almost no effect in the magnetic contribution of the Mg-Fe compound, whereas it clearly quenches the magnetic contribution for the Mg-Co-Fe compound above $T_{SG}$.

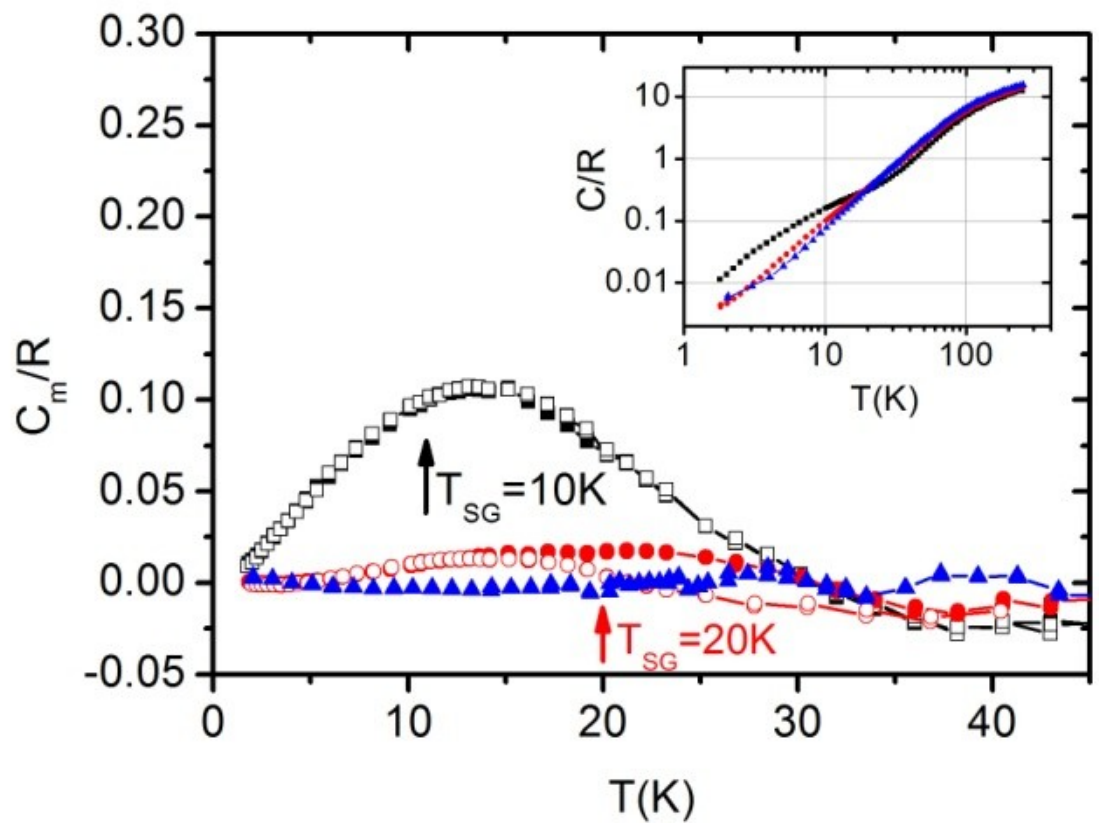

FIG. SM2. (Color on line) Magnetic contribution to the heat capacity of $MgFeBO_4$ (black squares), $Mg_{0.5}Co_{0.5}FeBO_4$ (red circles) and $CoFeBO_4$ (blue triangles) obtained at 0 (solid symbols) and 90 kOe (open symbols) from 2K to 45 K. Inset shows the total heat capacity in the whole temperature range, from 2 to 300 K.

---